\documentclass[journal]{IEEEtran}
\usepackage{pifont}
\usepackage{array}
\usepackage{booktabs}
\usepackage{multirow}
\usepackage{subfigure}
\usepackage[justification=centering]{caption}
\usepackage{times,amsmath,color,amssymb,amsfonts,graphicx,epsfig,cite,psfrag,subfigure,algorithm,epstopdf}
\usepackage{algorithmic}
\usepackage{textcomp}

\usepackage{bbding}

\def\BibTeX{{\rm B\kern-.05em{\sc i\kern-.025em b}\kern-.08em
    T\kern-.1667em\lower.7ex\hbox{E}\kern-.125emX}}

\begin{document}


\def\FIGDIR{./figs}

\IEEEoverridecommandlockouts

\title{Wireless and Service Allocation for Mobile Computation Offloading with Task Deadlines}


\author{\IEEEauthorblockN{Hong Chen\IEEEauthorrefmark{1}, Terence D.\ Todd\IEEEauthorrefmark{1}, Dongmei Zhao\IEEEauthorrefmark{1} and George Karakostas\IEEEauthorrefmark{2}\\}
\IEEEauthorblockA{\IEEEauthorrefmark{1}Department of Electrical and Computer Engineering \\
\IEEEauthorrefmark{2}Department of Computing and Software\\
McMaster University\\ Hamilton, Ontario, CANADA\\ Email:
  \texttt{\{chenh151,todd,dzhao,karakos\}@mcmaster.ca}\\[1ex]}
\thanks{This work has been submitted to the IEEE for possible publication. Copyright may be transferred without notice, after which this version may no longer be accessible.}
}

\maketitle \thispagestyle{empty}

\begin{abstract}

In mobile computation offloading (MCO), mobile devices (MDs) can choose to either execute tasks locally or to have them executed on a remote edge server (ES). This paper addresses the problem of assigning both the wireless communication bandwidth needed, along with the ES capacity that is used for the task execution, so that task completion time constraints are satisfied. The objective is to obtain these allocations so that the average power consumption of the mobile devices is minimized, subject to a cost budget constraint. The paper includes contributions for both soft and hard task completion deadline constraints. The problems are first formulated as mixed integer nonlinear programs (MINLPs). Approximate solutions are then obtained by decomposing the problems into a collection of convex subproblems that can be efficiently solved. Results are presented that demonstrate the quality of the proposed solutions, which can achieve near optimum performance over a wide range of system parameters.


\end{abstract}


\sloppy
\allowdisplaybreaks

\section{Introduction}
\label{sec:introduction}

Mobile computation offloading (MCO) can be used to improve mobile device (MD) performance by running computational tasks on a remote cloud server rather than executing them locally \cite{noor2018mobile,kwon2016precise,ba2013mobile}. Since the energy needed for task execution is incurred by the cloud server, a reduction in mobile device energy consumption can often be obtained \cite{gu2019real,zhang2017energy,VirtualCCProviderMobile,ElasticExecutionBetweenMobileCloud,3b,3d,Mazza2017}.  During MCO, wireless communications is used by the MD to communicate with the cloud server. This interaction incurs MD energy use that would not otherwise exist if the task were executed at the MD. MCO also incurs added latency due to the time needed for the MD to interact with the cloud server \cite{Q6,Q7}. An \emph{edge server} (ES) located close to the network base stations is typically used to reduce this delay by providing high interconnection bandwidth between the base station (BS) and the ES \cite{Huawei5G}.

The question of whether a given task should be offloaded has been studied extensively \cite{OptimizationRadioComputationalResourcesEnergyEfficiencyLatencyConstrained,Dab2019,Sheng2020,chen2020,Du2018,Yang2019,Zhang2018,R3,R4,Q5}. It is clear from this work that in order to obtain good performance, the offloading decisions should incorporate both the limited edge server computational capacity \cite{R3,R4,Q5}, and the temporal evolution of the system during the computation offload. This includes the queueing behaviour experienced by offloaded tasks awaiting execution at the ES \cite{Du2018,Yang2019,Zhang2018}. Prior work has also considered the question of how best to configure system resources so that MCO is best accommodated \cite{OptimizationRadioComputationalResourcesEnergyEfficiencyLatencyConstrained,Du2018,Yang2019,Zhang2018,R1,R2}. These are the issues that are considered in our paper and involve the tradeoffs between wireless communication and edge server capacity assignment and how these affect the delay performance experienced by the MDs.



The wireless and execution capacity assignment problem in MCO can be informally stated as follows. A network leaseholder (NL) purchases both wireless channel capacity and edge server execution services, subject to a cost budget constraint. The leased resources are then used to provide MCO to a large set of mobile devices \cite{Q4}. When an MD generates a task for execution, there is an associated deadline, which gives the time by which task execution should be completed with a high degree of certainty \cite{12}.  The objective is to find a joint wireless and ES resource assignment that minimizes the mean MD power consumption subject to the budget constraint and constraints on the task completion times.  Note that this problem is different than that of network slice creation \cite{intro3}. In this case, the NL simply purchases services from the network owner (NO), who prices the cost of unit wireless channel and computational resources. Due to the edge server placement, we consider the case where the dominant latencies are that of wireless access and edge server execution~\cite{Huawei5G}.

The paper is novel in that it includes formulations for both \emph{soft and hard task completion time deadlines}. In the soft deadline case, the wireless and edge server capacity assignments are designed so that the probability of task completion time deadline violation is upper bounded. In the hard deadline case, task execution deadlines must {\em always} be respected, which is accomplished by including concurrent local execution (CLE) \cite{Arvin} into the problem formulation. In CLE, local execution of the task may be initiated while offloading is ongoing, so that the task completion time deadline is always met.

The inclusion of task deadline constraints significantly increases the difficulty of the problem compared to that of prior work with no completion time requirements or that uses a mean delay criterion \cite{meandelay,Q2}. In order to obtain solutions to the problem, a queuing model is used to obtain the delay distribution experienced by tasks that are offloaded to the ES \cite{Q1,Q2}.  This model is incorporated into the resulting optimization problems, which are formulated as mixed integer nonlinear programming problems (MINLPs) that are computationally hard to solve exactly. Approximate solutions are obtained by decomposing the non-convex non-linear formulation into a collection of convex subproblems that can be solved efficiently, and then picking the best of these solutions.

A variety of results are presented that characterize the tradeoffs between task deadline violation, average MD power consumption and the cost budget. Our results show the quality of the proposed solutions, which can achieve close-to-optimum performance for a wide range of system parameters. The results also show that with CLE, the proposed solution not only guarantees respecting {\em all} hard task completion deadlines, but does so with only slightly higher MD power consumption when compared to the soft task completion deadlines solution with a small deadline violation probability. On the other hand, we show that there is an apparent trade-off in the case of soft task completion deadlines between the average power consumption and the deadline violation probability. Namely, the average MD power consumption of our solution is significantly reduced when a higher deadline violation probability is tolerable.

The main contributions of the paper are summarized below.
\begin{itemize}

\item This paper addresses the problem of assigning computational and wireless channel resources for MCO, subject to task execution completion time deadlines. The work is the first that generates joint resource assignments for both soft and hard task deadlines using very general system modelling assumptions compared to prior work. The soft deadline case aims to create assignments so that the probability of task completion time deadline violation is upper bounded. In the hard deadline case, the paper is also unique in that it creates resource assignments where task completion time deadlines are always satisfied. This is done by incorporating CLE into the problem formulations. For this reason, this is the first paper that obtains system resource assignments for MCO that ensure that task completion time deadlines are always satisfied.

\item Modeling both soft and hard job completion time targets significantly increases the difficulty of the problem compared to prior work with no completion time requirements or that uses a mean delay criterion [30] [31]. In both deadline cases, the paper addresses this by incorporating an ES queueing system into the problem formulation that models the delay distribution experienced by arriving tasks. The assignment problem is addressed by inverting the estimated probability density function (PDF) of the task completion time and incorporating it into the optimizations. These resource assignments are obtained under very general modeling assumptions, where the wireless channels are modeled as arbitrary base station specific sets of Markov processes and task execution times have a general probability distribution.

\item The problems are first formulated as MINLPs, with {\em integral} decision variables for the number of wireless channels reserved, and a {\em continuous} decision variable for the portion of ES reserved. Even the relaxations of these MINLPs are difficult to solve, since they are non-convex. Hence, instead of following the common practice of solving the relaxation and rounding the fractional solution, we observe that the discretization of the {\em continuous} variable and the replacement of the {\em discrete} channel variables by approximate functions of the {\em continuous} blocking probabilities, allows us to break the original non-convex MINLPs into collections of {\em convex} subproblems, that can be solved efficiently. Our solutions are approximate, and their accuracy depends on both the discretization granularity and the approximation functions used for blocking probabilities. On the other hand, they are based on very general assumptions, i.e., the existence of convex upper bound approximations of the inversion of blocking probabilities. The more restricted the system model is, the better these approximations are.

\end{itemize}

The remainder of the paper is organized as follows. In Section \ref{sec:relatedwork} the prior work most related to our paper is reviewed. The system model and problem formulation is then described in Section \ref{sec:systemmodel}. In Section \ref{sec:softdeadlines}, the general design problem is first considered assuming soft task completion time deadlines, where the probability of deadline violation is bounded. Following this, in Section \ref{sec:harddeadlines} a formulation is described when task completion times are subject to hard deadlines. The problem formulations in both cases are non-convex and difficult to deal with directly using conventional optimization approaches. In Section \ref{sec:approximatesolutions}, approximation solutions are proposed where the original problems are decomposed into convex subproblems that can be efficiently solved. Both the soft and hard deadline cases are considered in Sections \ref{sec:approximatesoftsolution} and \ref{sec:approximateharddeadlines}. Section \ref{sec:poissonassumptions} then introduces some common system assumptions used in the remainder of the paper when solving the optimizations. Both the soft and hard deadline cases are then treated in detail in Sections \ref{sec:poissonapproximatesoftdeadlines} and \ref{sec:poissonapproximateharddeadlines}. In Section \ref{sec:simulationresults} simulation results that demonstrate the proposed designs are given. Both the single class and multiple classes of tasks cases are considered in Sections \ref{sec:simulationresultssingle} and \ref{sec:simulationresultsmultiple}. Finally, we present our conclusions of the work in Section \ref{sec:conclusions}.

\section{Related Work}
\label{sec:relatedwork}

A large amount of prior MCO work considers the problem based on system state inputs sampled at task generation times, i.e., the models assume that the system is static throughout the offload period \cite{OptimizationRadioComputationalResourcesEnergyEfficiencyLatencyConstrained,Du2018,Yang2019,Zhang2018,R1,R2,R3,R4,Dab2019,chen2020,R7,R8,Q5}. As in our paper, task offloading decisions become more complex when the MD interacts with the network over wireless channels that may change randomly during the offload. Reference \cite{Q1} studies a distributed computation offloading problem with delay constraints using stochastic communication channels but does not take into account the energy consumption incurred during task offloading.  The work in \cite{meandelay} uses a Markov decision process that analyzes the mean task delay and the average system throughput.  Unlike our paper, a throughput maximization problem is formulated with constraints on the average task delay, rather than using the delay distribution.  In \cite{Q2}, task offloading is modeled as a game using a network of queues to obtain the end-to-end delay. The problem is transformed into one with a generalized Nash equilibrium solution that captures the conflicting interests in resource allocation among mobile network operators and computing resource providers. In references \cite{meandelay} and \cite{Q2} the average delay is considered rather than the stringent types of soft and hard delay constraints considered in our paper.
Reference \cite{Q3} considers task offloading with statistical QoS guarantees (i.e., tasks are allowed to complete before a given deadline with a probability above a given threshold) to maximize the MD energy efficiency. The energy efficiency is defined as the ratio of the overall executed (transmitted) bits of tasks to the total energy consumption of the MDs. Statistical computation and transmission models are introduced to quantify the correlation between the statistical quality of service (QoS) guarantee and task offloading process. Unlike the models used in \cite{Q2} and \cite{Q3}, our paper uses a task offloading and resource allocation formulation that uses very general system model assumptions, including base station specific sets of Markov processes for channel modelling.

\begin{table*}[htbp]
\begin{center}
\caption{Related Work Summary}
\label{RelatedWorkTable}
\begin{tabular}{|c|m{2.5cm}<{\centering}|m{2cm}<{\centering}|m{2cm}<{\centering}|m{2cm}<{\centering}|m{2cm}<{\centering}|}
\hline
References  & Joint channel and computation resource assignment & Soft task deadlines & Hard task deadlines &  Resource expense & Temporal evolution \\ \hline
\cite{chen2020}\cite{R3}\cite{R4}\cite{Q5} &   &   & $\checkmark$ &    &  \\ \hline
\cite{R1}\cite{R9} & \checkmark &   &   &    &  \\ \hline
\cite{Q4}  &   &   & $\checkmark$ & $\checkmark$  &  \\ \hline
\cite{meandelay}\cite{Q2}  & \checkmark &   &    &    & \checkmark \\ \hline
\cite{Q1}  &   &   & $\checkmark$  &    & \checkmark \\ \hline
\cite{Q3}  & \checkmark & \checkmark &    &    & \checkmark \\ \hline
Our paper  & \checkmark & \checkmark & \checkmark  & \checkmark & \checkmark \\ \hline
\end{tabular}
\end{center}
\end{table*}

Reducing both mobile energy consumption and task execution time is a common objective in mobile computation offloading. The work in \cite{R9} investigates a latency minimization problem in a multi-user time-division multiple access system with joint communication and computation resource allocation. Our paper, instead, uses a soft task deadline criterion based on modelling the distributions of both upload and execution time delays. Hard completion time constraints are considered in references \cite{chen2020,R3,R4,R7,Q1,Q5}. However, unlike our work, they consider the hard completion time requirement as a constraint in the problem formulation. For this reason, if the provided network resources or the MD transmit power are insufficient, the hard completion time constraints may not be satisfied. In our work, we avoid this infeasibility by applying CLE that ensures that hard completion time constraints are always satisfied. A benefit from integrating CLE into the problem formulation is that we no longer require the hard completion time constraints in the problem formulation. The objective in \cite{R3} is to minimize the energy consumption of the entire system, and in references \cite{R4} and \cite{R7}, the objective is to minimize the total energy consumption of all MDs. Instead of satisfying delay constraints, the work in \cite{R1,R2,R8} optimize a utility function that is a weighted sum of task completion time and energy consumption. Unlike the above work, two different kinds of delay constraints are introduced in our paper, i.e., soft deadlines captured by the statistics of the completion time of the tasks and hard deadlines that are always satisfied by CLE.

Prior work has considered the optimization of wireless network and computational server resources to improve MCO performance \cite{OptimizationRadioComputationalResourcesEnergyEfficiencyLatencyConstrained,Du2018,Yang2019,Zhang2018,R1,R2}. In particular, offloading decisions and base station associations are optimized with transmission power and channel assignments in a cellular network to minimize the total energy consumption of all MDs, subject to task's latency constraints \cite{chen2020}. Reference \cite{R3} studies the problem of task offloading and channel resource allocation for ultra-dense networks and minimizes the total energy consumption of the system with a limited delay tolerance. The work in \cite{R4} studies MCO by considering application latency fairness and minimizes MD energy consumption by jointly optimizing the offloading ratio, channel assignments, and channel time allocations. Reference \cite{Q5} investigates the power minimization problem for meeting the service delay requirements in multi-cell multi-user mobile edge computing networks. Channel assignment and power allocation problems are considered jointly. The work in \cite{Q4} studies the joint resource management of link scheduling, channel assignment and power control for device-to-device communication assisted multi-tier fog computing with the objective of maximizing the network operator profit under deadline requirements. It considers the service charge collected from all end users, total expense in renting third-party fog nodes, and the energy cost of the ES. All of this work \cite{R3,R4,chen2020,Q5,Q4} optimizes radio resources and offloading decisions without considering edge server computational capability. The work in \cite{R1} investigates relay-assisted computation offloading to minimize the weighted sum of task execution delay and the energy consumption by jointly optimizing the offloading ratio, bandwidth allocation, processor speeds, and transmit power.



Table \ref{RelatedWorkTable} summarizes the work described above that is most related to our paper, and compares it to this paper on five key properties:
\begin{LaTeXdescription}
\item[\bf Joint channel and computation resource assignment:] The column denotes work where both channel and computation resource assignments are jointly generated. Our work differs from the rest in that we assign aggregate channel resources from the network operator to each base station so that it can support its associated mobile device population, i.e., we do not allocate  channel and computation resources of each BS and ES to individual MDs.

\item[\bf Soft task deadlines:] The work selected in this column considers some form of soft (i.e., statistical) task deadlines. However, the models we use in this paper are quite different with more general underlying assumptions.  Since our soft deadline model aims to set bounds on the probability of task deadline violation, we model the complete delay distribution experienced by executed tasks. This includes the base station channel delay (which is modeled by base station specific Markov processes) and the queueing delay experienced at the ES, where execution times can have a general distribution.

\item[\bf Hard task deadlines:] Although there is other work selected in this column, a significant difference exists compared with our paper, which we have already discussed above. Namely, our work can {\em always} satisfy all hard task deadlines by incorporating the CLE mechanism into the modeled system. The related work, instead, considers the existence of hard deadlines as a problem constraint that may result in problem infeasibility, which can never happen in our case.

\item[\bf Resource expense:] This column denotes work where the resources provided to the MDs are charged by a third-party (e.g., network operator). The work selected considers computational resource expense but not on the wireless base station side. A network profit maximization problem is studied where an expense budget is not considered, unlike the case in our work.

\item[\bf Temporal evolution:] Temporal evolution means that the offload periods may include stochastic changes to the wireless channels and the ES, so that this information must be modeled in the problem formulation, as in our paper. The randomness modeled in the selected work has different underlying assumptions compared to our paper.
\end{LaTeXdescription}



\section{System Model and Problem Formulation}
\label{sec:systemmodel}

\begin{figure}[t]
  \centering
  \includegraphics[width=85mm]{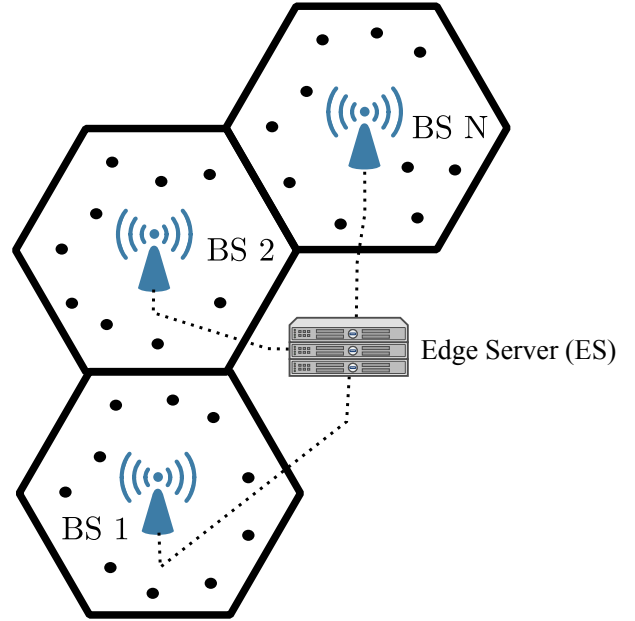}
  \caption{System Model}
  \label{fig:1}
\end{figure}

As shown in Fig.~\ref{fig:1}, we consider a network that consists of $N$ BSs that are owned and operated by a NO. The set of BSs is denoted by $\mathcal{N}=\{1,2,\ldots,N\}$ and indexed by $n\in\mathcal{N}$. The network also contains an ES. Tasks generated by an MD can be offloaded through the wireless network and executed on the ES.

The NO permits a NL to rent wireless communication and ES computational capacity that the NL can use for mobile computation offloading for its MDs.  When this is done, for each BS $n$, there are up to $K_n$ available channels that can be selected by the NL. The cost of renting a channel from BS $n$ is set by the NO to $\alpha_n$.
When a channel is included in the agreement, the NO agrees to provision its network so that sufficient resources are available to allow the traffic generated on the channel to be carried to the ES with an acceptable delay with a high degree of certainty. Since the ES is located at the edge of the network, we focus on the dominant sources of delay, i.e., wireless access at the BSs and task execution at the ES~\cite{Huawei5G}.

In order to use the computing resources at the ES, the NL must also lease CPU resources at the ES. The cost (based on the number of CPU cycles per second) for leasing on the CPU resource is denoted by $\beta$. The maximum available CPU speed for rental is $f^{\rm{C}}$ CPU cycles per second.

When an agreement is made between the NO and NL, $x_n$ is defined as the number of channels from BS $n$ that are included, and $y \in \left[ {0,1} \right]$ is defined as the fraction of maximum CPU speed at the ES that is included, i.e., the CPU speed available for the NL will be $y f^{\rm{C}}$. It is assumed that the NL has a cost budget, denoted by $B^{\text{max}}$. Accordingly, the total rent must satisfy the following constraint:
\begin{equation} \label{eq:budget}
\textstyle\sum_{n= 1}^N {{\alpha _n}{x_n}}  + \beta y{f^{\rm{C}}}  \le {B^{\max }}.
\end{equation}

There are $J$ classes of tasks generated by the MDs, which may need to be offloaded to the ES. Let $\mathcal{J}=\{1,2,\ldots,J\}$ be the set of task classes. The class $j$ of a task is defined by parameters $s_j$, $q_j$, and $d_j$, where $s_j$ is the input data size in bits, $q_j$ is the computation load in number of CPU cycles, and $d_j$ is the deadline of the task in seconds. In what follows, $\tilde d_j=\left\lfloor {d_j/\tau} \right\rfloor$ is the task deadline rounded down to time slots of the same duration $\tau$ as the wireless transmission time slots (see below). The probability of a task generated by an MD belonging to class $j$ is denoted by $P_j^{\rm{C}}$; we assume that this probability is known, e.g., by observing the past history of offloading requests.

Our objective is to create a NO/NL contract for MCO. In MCO, tasks generated by an MD can be executed either locally (at the MD itself) or offloaded through the network and executed on the ES. We focus on two goals, each depending on how {\em hard} the task deadline constraint is. Our first goal is to accomplish this so that the mean mobile power consumption is minimized subject to the cost budget constraint and such that the probability that task execution deadline violation is bounded, i.e., the deadline constraints can be violated, albeit rarely. Our second goal is to create a power-efficient, budget-respecting assignment which respects {\em all} task deadlines, i.e., deadline constraints are hard; for that purpose we will employ CLE \cite{Arvin}.

We model the wireless channels between the MDs and the BSs as discrete-time Markov processes. It is assumed that there are $I_n$ channel models for BS $n$, which are a function of the radio propagation environment that the MDs experience at that BS. $\mathcal{I}_n=\{1,2,\ldots,I_n\}$ is the set of all  wireless channel models in BS $n$. For each of the channel models, the Markovian transition probabilities are defined in the usual way, i.e., given the channel state in the current time slot, there is a probability associated to its transition to another state in the next time slot. The time slot duration is defined to be $\tau$ seconds. A class $j$ task, offloaded to BS $n$ by the MD, encounters channel model $k$ with probability $P^{\rm{G}}_{n,j,k}$; as with task generation probabilities $P_j^{\rm{C}}$ above, we assume that this probability is also known, e.g., by observing the past history of offloading requests.

To obtain the design, the decision to offload the execution of a task is made using a \emph{local execute on blocking} (LEB) mechanism as follows. When an MD in BS $n$ generates a class $j$ task, the MD offloads the task if at least one of the $x_n$ channels is available for immediate use. Otherwise, the MD executes the task locally. When a channel is available, the MD begins the offload by uploading the $s_j$ task bits needed for execution on the ES. The LEB mechanism is useful in that either local execution or remote offloading is initiated immediately at task release time, which may be advantageous when task deadlines are tight. It also provides a simple mechanism for assessing when the  current level of local congestion is high, which would suggest that local execution is beneficial.



Tasks arrive at BS $n$ according to a stationary process with average arrival rate $\lambda_n$ tasks per second. According to the LEB mechanism, a new task is blocked from BS channel access if all the $x_n$ channels are busy with uploading other tasks. We denote the task blocking probability at BS $n$ by $P_{{\rm{B}}n}(x_n)$, which is a function of $x_n$. For the sake of notation simplicity, we use $P_{{\rm{B}}n}$ in the rest of the paper. Let $p^\text{L}$ be the power needed in the MD to process tasks. When a class $j$ task is blocked from offloading and executed locally, the local execution time is given as
 $L_j = {q_j} / f$,
where $f$ is the MD's execution speed in number of CPU cycles per time slot\footnote{$L_j$ is normally measured in CPU cycles, but in order to apply CLE and to simplify the system, we round it up to a multiple of $\tau$.}.
Define $\bar L$ as the average local execution time of tasks. Since the task blocking is caused by channel access, which is the same for all task classes, we have $\bar L =\sum_{j = 1}^J P_j^{\rm{C}}L_j$. The average energy consumption for executing a task locally is given by $p^{\rm{L}} {\bar L}$. Consider all the tasks that are generated in BS $n$ and blocked from offloading in one second, then the mean energy for executing these tasks locally is
\begin{equation}
\label{eq:EiL}
E_{n}^{\rm{L}}(x_n) = {{P_{{\rm{B}}n}}{{\lambda _n}{p^{\rm{L}}}{\bar L}} },
\end{equation}
which is the average power consumption of the MDs.


The wireless upload transmission time $t_{n,j,k}^{\text{W}}$ of a $j$th class task in BS $n$ when the wireless channel model is $k$, is measured in time slots. The mean wireless upload transmission time $\bar t^{\rm{W}}_{n,j,k}$ for $j$th class tasks in BS $n$ according to channel model $k$ can be calculated, since $\Pr [ {t_{n,j,k}^{\rm{W}} = l}]$ can be computed for all $l$ from channel model $k$.
Moreover, the mean wireless transmission time $\bar t_n^{\rm{W}}$ for BS $n$ is
\begin{equation}
\bar t_n^{\rm{W}}=\sum\limits_{j = 1}^J \sum\limits_{k = 1}^{{I_n}} {P_j^{\rm{C}}}P^{\rm{G}}_{n,j,k} \bar t_{n,j,k}^{\rm{W}} .
\end{equation}


Under the stated assumptions, the aggregate mean task arrival rate $\lambda$ at the ES is given by
\begin{equation}  \label{eq:aggregate}
\lambda  = \textstyle\sum_{n = 1}^N {( {1 - {P_{{\rm{B}}n}}} ) {{\lambda _n}} } .
\end{equation}
As is normally the case for stability in a single server queueing system, the following constraint must always be satisfied,
\begin{equation}
\label{eq:stability}
\lambda < {\mu ^{\rm{C}}},
\end{equation}
where ${\mu ^{\rm{C}}}$ denotes the mean service rate at the ES, i.e, ${\mu ^{\rm{C}}}= yf^{\rm{C}}/\sum_{j = 1}^J P_j^{\rm{C}}q_j$. As will become clear later, we can relax this constraint to
$\lambda \leq yf^{\rm{C}}/\sum_{j = 1}^J P_j^{\rm{C}}q_j $
without affecting our proposed solutions.

Let $t_{n,j,k}^{\rm{C}}$ be the delay (including both queueing and execution time) experienced by a $j$th class task from BS $n$ at the ES, under wireless channel model $k$. It takes continuous values, and $\Pr [ {{t_{n,j,k}^{\rm{C}}} \le t } ]$, for any $t\geq 0$, is a function of $\lambda$ and $\mu ^{\rm{C}}$.
In what follows, ${\tilde t}_{n,j,k}^{\rm{C}}$ is the discretization of $t_{n,j,k}^{\rm{C}}$, measured in time slots; its distribution is calculated by
\begin{align}
\label{eq:tB}
\Pr[ {{{\tilde t}_{n,j,k}^{\rm{C}}} = b}] =\Pr[ {{t_{n,j,k}^{\rm{C}}} \le b\tau }] - \Pr[ {{t_{n,j,k}^{\rm{C}}} \le (b - 1)\tau}]
\end{align}
for any number of time slots $b\geq 0$. Table~\ref{Notations} lists the related notation and their associated meanings.


\renewcommand\arraystretch{1.4}
\begin{table}[htbp]
\begin{center}
\caption{Summary of Notation}
\label{Notations}
\begin{tabular}{|m{0.09\columnwidth}|m{0.55\columnwidth}|m{0.2\columnwidth}|} 
\hline
Notation     &     Definition & Units \\
\hline
$\mathcal{N}$  &  Set of BSs, $|\mathcal{N}|=N$ &\\
\hline
$\mathcal{J}$ &  Set of task classes, $|\mathcal{J}|=J$  &\\
\hline
$\mathcal{I}_n$ & Set of channel models of BS $n$, $|\mathcal{I}_n|=I_n$ & \\
\hline
$K_n$   &  Number of available channels in BS $n$ &\\
\hline
$f^{\rm{C}}$ &  Maximum available ES capacity & CPU cycles/sec\\
\hline
$\alpha_n$ & Unit price of wireless channels from BS $n$ & \$ per channel \\
\hline
$\beta$ &  Unit price of ES capacity & \$ per bps\\
\hline
$x_n$ &  Number of channels from BS $n$  & \\
\hline
$y$ &  Fraction of maximum ES capacity  &\\
\hline
$B^{\max}$ & Cost budget & \$ \\
\hline
$s_j$ & Data size of a task in class $j$  & bits\\
\hline
$q_j$ & Computation load of a task in class $j$  & CPU cycles\\
\hline
$d_j$ & Deadline of a task in class $j$  & sec\\
\hline
$\tilde d_j$ & Discretized deadline of a task in class $j$ & Time slots \\
\hline
$P_j^{\rm{C}}$ & Probability of a task belonging to class $j$ & \\
\hline
$P^{\rm{G}}_{n,j,k}$ & Probability of a class $j$ task in BS $n$ with channel model $k$  &\\
\hline
$P_{{\rm{B}}n}$ & Blocking probability in BS $n$  &\\
\hline
$\mu^{\rm{C}}$ & Mean service rate at the ES & Tasks/sec \\
\hline
$\lambda_n$ & Average task arrival rate in BS $n$ & Tasks/sec  \\
\hline
$\lambda$ & Aggregate average task arrival rate at ES  & Tasks/sec\\
\hline
$\tau$   & Time slot & sec \\
\hline
$t_{n,j,k}^{\text{W}}$ & Wireless transmission time of a $j$th class task in BS $n$ with channel model $k$ & Time slots\\
\hline
$\bar t_{n,j,k}^{\text{W}}$ & Mean wireless transmission time of a $j$th class task in BS $n$ with channel model $k$ & Time slots\\
\hline
$\bar t_{n}^{\text{W}}$ &  Mean task uploading transmission time in BS $n$ & Time slots \\
\hline
$t_{n,j,k}^{\rm{C}}$ &  Execution time at ES for class $j$ tasks from BS $n$ with channel model $k$ & sec  \\
\hline
${\tilde t}_{n,j,k}^{\rm{C}}$ &  Discretized value of $t_{n,j,k}^{\rm{C}}$ & Time slots \\
\hline
$t_j^{\rm{L}}$ &  Latest feasible starting time for local execution  & Time slots\\
\hline
$\varepsilon_{j}$ &   Tolerable probability a class $j$ task exceeds deadline & \\
\hline
$p^{\rm{L}}$ &  Local energy consumption per time slot  & Joules\\
\hline
$p^{\rm{T}}$ &  Wireless transmission energy per time slot & Joules \\
\hline
$E_n^{\rm{T}}$ & Average MD power consumption for uploading tasks in BS $n$ & Watts \\
\hline
$E_n^{\rm{C}}$ & Average MD power consumption for uploading and executing tasks in BS $n$  & Watts \\
\hline
\end{tabular}
\end{center}
\end{table}

\subsection{Problem Formulation with Soft Deadlines}
\label{sec:softdeadlines}

We consider the distribution of total delay for an offloaded task, which is the sum of the {\em data upload} delay $t_{n,j,k}^{\rm{W}}$ and the {\em task execution at ES} delay $t_{n,j,k}^{\rm{C}}$, for BS $n$, task class $j$, and channel model $k$. Note that both delays are random variables. As mentioned earlier, the data transmission delay from the BS to the ES is negligible. In addition, in this paper we consider the case of a very small amount of data returned once the execution is completed, and, therefore, we consider only uploading delays between MD and BS.

Following common practice (e.g., \cite{12}) in modelling soft deadlines along the lines of QoS requirements, a $j$th class task in BS $n$ under wireless channel model $k$, must have a total delay satisfying
\begin{equation}   \label{eq:totald}
{\rm{Pr}}[ {t_{n,j,k}^{\rm{W}}  + t_{n,j,k}^{\rm{C}} \le {d_j}}] \ge 1 - {\varepsilon _{j}},
\end{equation}
where $0< \varepsilon_{j} \le 1$ is the (given) tolerated probability the completion time of a class $j$ task exceeds its deadline.\footnote{The case $\varepsilon_j=0$ corresponds to the case of hard deadlines, and will be dealt with in the next section.} Note that $ t_{n,j,k}^{\rm{W}}$ takes discrete values (number of time slots), $t_{n,j,k}^{\rm{C}}$ takes discrete values (number of CPU cycle periods), while $d_j$ is continuous (in seconds), so \eqref{eq:totald} assumes that all quantities are first converted to secs. Its LHS is a function of $x_n,y$.

The joint probability distribution of total delay is
\begin{multline}  \label{Eq:d}
\Pr[ {t_{n,j,k}^{\rm{W}}  + t_{n,j,k}^{\rm{C}} \le {d_j}}] = \\
\sum_{l = 1}^{l_{\max}} {\Pr [ {t_{n,j,k}^{\rm{W}} = l } ]}  \Pr [ {t_{n,j,k}^{\rm{C}} \le{d_j} - l\tau } ] ,
\end{multline}
where $l_{\max}=\lfloor (d_j - {q_j}/{y f^{\rm C}}) /{\tau} \rfloor $ is the maximum value that $l$ can take, since ${q_j}/{y f^{\rm C}}$ is the execution time at the ES without queueing.

The average power consumption of MDs in BS $n$ to upload tasks that are granted channels for offloading is
\begin{equation}  \label{eq:trans}
E_{n}^{\rm{T}}(x_n) = {( {1 - {P_{{\rm{B}}n}}} ) {{\lambda _n}} {p^{\rm{T}}}\bar t_n^{\rm{W}}},
\end{equation}
where ${p^{\rm{T}}}$ is the transmission energy per time slot used by the MD for uploading the task bits. Therefore, the expected average power consumption of the MDs for uploading and executing tasks arriving at BS $n$ is $E_{n}^{\rm{L}}(x_n) + E_{n}^{\rm{T}}(x_n)$.

Our objective is to create an allocation that minimizes $E_{n}^{\rm{L}}(x_n) + E_{n}^{\rm{T}}(x_n)$ under the cost budget and deadline constraints \eqref{eq:budget} and \eqref{eq:totald}. The problem can be formulated as follows:
\begin{alignat}{2}     \label{Eq:minGE}
\mathop{\min}_{\bold{x}, y } \sum\limits_{n = 1}^N
 [E_{n}^{\rm{L}}(x_n) + E_{n}^{\rm{T}}&(x_n)]  \text{ s.t.} \\
\sum\limits_{n = 1}^N {{\alpha _n}{x_n}}  + \beta {f^{\rm{C}}} y  &\le {B^{\max }} \label{Eq:1C2} \\
{\rm{Pr}}[ {t_{n,j,k}^{\rm{W}}  + t_{n,j,k}^{\rm{C}} \le {d_j}} ] &\ge 1 - {\varepsilon_{j}}, &&\forall  n,j,k
\label{Eq:1C3}\\
(f^{\rm{C}}/\sum\limits_{j = 1}^J P_j^{\rm{C}}q_j) y &\ge \lambda  \label{Eq:1C6} \\
x_n &\in \{0,1,\ldots,K_n\},\ \ \   &&\forall  n\in\mathcal{N} \label{Eq:1C4}\\
0 \le y &\le 1 \label{Eq:1C5}.
\end{alignat}
Constraints \eqref{Eq:1C2} and~\eqref{Eq:1C3} are constraints \eqref{eq:budget} and \eqref{eq:totald}. Constraint~\eqref{Eq:1C6} is the (relaxed) queue stability requirement for ES; it is equivalent to~\eqref{eq:stability}, since equality leads to infinite mean queueing delay, which is never optimal. The optimization problem \eqref{Eq:minGE}-\eqref{Eq:1C5} is a mixed integer nonlinear programming (MINLP) problem.
Constraint~\eqref{Eq:1C4} ensures that the number of channels assigned does not exceed the maximum number available in each BS.
Even the fractional relaxation of MINLP problem \eqref{Eq:minGE}-\eqref{Eq:1C5} is non-convex, due to its objective and constraints \eqref{Eq:1C3}, and, as a result, it is computationally inefficient to solve it exactly. Hence we are going to propose approximate solutions for it.

\subsection{Problem Formulation with Hard Deadlines}
\label{sec:harddeadlines}

For the case of {\em hard} deadline constraints, i.e., when the task deadline {\em must} be respected, we employ CLE \cite{Arvin}. In CLE, local execution of the task may be initiated while offloading is
ongoing, so that the task deadline is always met, even if offloading fails to finish in time due to the stochastic nature of the wireless channels.
Guaranteeing task completion before its deadline may incur additional costs (due to potentially simultaneous
local and remote execution of the same task).

When CLE is employed, and in order to ensure that the local execution of a task from class $j$ finishes by its deadline,
the latest feasible starting time for local execution is
\begin{equation}
t_j^{\rm{L}}=\tilde d_j-L_j+1.
\end{equation}

The expected wireless transmission power is still given by \eqref{eq:trans}. However, due to the overlap of offloading and local execution because of CLE, there is an extra mean power consumption due to a (potential) overlap with local execution. This expected overlap power consumption is
\begin{align}   \label{eq:over}
&E_{n,j,k}^O (x_n,y) =  (1 - P_{\rm{B}n} ) \lambda _n  \nonumber \\
&\cdot \sum\limits_{t = t_j^{\rm{L}}}^{{{\tilde d}_j}}\!\! \sum\limits_{l = 1}^{t - \lceil {\frac{q_j}{y{f^{\rm C}}\tau}} \rceil } \!\!\!\! \Pr [ t_{n,j,k}^{\rm{W}} = l  ]\Pr [ {{{\tilde t}_{n,j,k}^{\rm{C}}}= t - l } ] \cdot {p^{\rm{L}}}( {t - t_j^{\rm{L}} + 1} ) ,
\end{align}
where $t$ is the number of time slots needed to complete the offloaded task, and $( {t - t_j^{\rm{L}} + 1} )$ is the offloading and local execution overlap. Note that $\Pr [ {{{\tilde t}_{n,j,k}^{\rm{C}}}= t - l } ]$ is a function of $x_n$ and $y$.

In case the task offloading goes beyond the finish of the local execution of a task, there is an extra power consumption incurred, whose expected value is
\begin{align}  \label{eq:waste}
&E_{n,j,k}^B (x_n,y)=  (1 - P_{\rm{B}n} ) \lambda _n \nonumber \\
&\cdot \sum\limits_{t = {{\tilde d}_j} + 1}^{ + \infty }\!\!\!\! {\sum\limits_{l = 1}^{t - \lceil {\frac{q_j}{y{f^{\rm C}}\tau}} \rceil } {\Pr [ {t_{n,j,k}^{\rm{W}} = l} ]\Pr [ {{{\tilde t}_{n,j,k}^{\rm{C}}} = t - l} ]} }{p^{\rm{L}}}{L_j} .
\end{align}
Hence, the expected power consumption of MDs for offloaded tasks in BS $n$ in one second is
\begin{align}\label{Eq:EC}
&E_n^{\rm{C}}(x_n,y) =  E_{n}^T(x_n) \nonumber\\
& +  \sum\limits_{j = 1}^J \sum\limits_{k = 1}^{{I_n}} {P_j^{\rm{C}}}P^{\rm{G}}_{n,j,k}  [E_{n,j,k}^O (x_n,y)+E_{n,j,k}^B (x_n,y) ],
\end{align}
and the expected power consumption of MDs for tasks arriving at BS $n$ in one second is $E_{n}^{\rm{L}}(x_n) + E_{n}^{\rm{C}}(x_n,y)$.

As before, our objective is to minimize the total expected power consumption of the MDs for uploading and executing the tasks that are generated in one second, but now subject to hard deadline constraints. The problem is formulated as follows:
\begin{align}    \label{Eq:minGCLE}
\mathop {\min } \limits_{\bold{x},y}  \sum\limits_{n = 1}^N[ E_{n}^{\rm{L}}(x_n) + E_{n}^{\rm{C}}&(x_n,y)] \text{ s.t.} \\
\sum\limits_{n = 1}^N {{\alpha _n}{x_n}}  + \beta {f^{\rm{C}}} y  &\le {B^{\max }}  \label{Eq:1CC2} \\
(f^{\rm{C}}/\sum\limits_{j = 1}^J P_j^{\rm{C}}q_j) y &\ge \lambda   \label{Eq:1CC5} \\
x_n &\in \{0,1,\ldots,K_n\}, \ \ \ \ \forall  n\in\mathcal{N}    \label{Eq:1CC1}\\
0 \le y &\le 1 \label{Eq:1CC4}.
\end{align}


\section{General Approximate Allocation Solutions}
\label{sec:approximatesolutions}

In this section, we propose approximate solutions for optimization problems \eqref{Eq:minGE}-\eqref{Eq:1C5} and \eqref{Eq:minGCLE}-\eqref{Eq:1CC4}, by decomposing them into convex optimization subproblems which can be solved efficiently.

\subsection{Approximate Solution for Soft Deadlines}
\label{sec:approximatesoftsolution}

In this subsection, we propose an approximate solution for the optimization problem
\eqref{Eq:minGE}-\eqref{Eq:1C5} by decomposing it into several convex
subproblems that can be solved efficiently, solve them, and then keep the best solution.
More specifically, we discretize variable $y\in [0,1]$
by breaking $[0,1]$ into $Y$ equal segments, so that $y$ takes values $y_a=a/Y$, for $a=0,1,\ldots,Y$.
With $y$ fixed, we show that the relaxation of \eqref{Eq:minGE}-\eqref{Eq:1C5} can be approximated
by a convex optimization problem, which can be solved in polynomial time.
The resulting (fractional) $x_n$'s are then rounded to integer values (and this is another source of suboptimality
for our solution method). After solving the resulting $Y+1$ problems, we output the minimum solution $x^*,y^*$.
Obviously, the quality of the approximation depends on the discretization parameter $Y$.

We consider the relaxed version of problem \eqref{Eq:minGE}-\eqref{Eq:1C5}, i.e., constraint
\eqref{Eq:1C4} has been replaced by $x_n\geq 0, \forall n$.
With $y$ fixed, we show that the non-convex problem \eqref{Eq:minGE}-\eqref{Eq:1C5} can be transformed into an
equivalent convex optimization problem with the $P_{{\rm{B}}n}$'s as the decision variables. First, we concentrate
on constraints \eqref{Eq:1C3}, \eqref{Eq:1C6}.
Note that ${\rm{Pr}}[ {t_{n,j,k}^{\rm{W}}  + t_{n,j,k}^{\rm{C}} \le {d_j}}]$
is a monotonically decreasing function of the aggregate mean task arrival rate $\lambda$.
Hence, by binary search in the range $[0,yf^{\rm{C}}/\sum_{j = 1}^J P_j^{\rm{C}}q_j]$, we can approximate within any desired accuracy
the maximum possible value of $\lambda$ that satisfies constraints \eqref{Eq:1C3} for all $n,j,k$. Let $\lambda^*$
be this maximum value (note that $\lambda^*<\mu^{\rm{C}}$, so stability is ensured). Using \eqref{eq:aggregate}, constraints \eqref{Eq:1C3}, \eqref{Eq:1C6} can be replaced by constraint
\begin{align}   \label{eq:lambda}
\sum_{n = 1}^N {( {1 - {P_{{\rm{B}}n}}} ) {{\lambda _n}} } \le {\lambda ^*}.
\end{align}

Next, we note that the blocking probability $P_{{\rm{B}}n}$ is monotonically decreasing in $x_n$; let
$P_{{\rm{B}}n}^{\text{min}}$ be the blocking probability when $x_n=K_n$. Then constraints \eqref{Eq:1C4}
can be replaced by the equivalent constraints
\begin{align}\label{Eq:Ineq}
P_{{\rm{B}}n}^{\text{min}} \le P_{{\rm{B}}n} \le 1, \ \forall n\in\mathcal{N}.
\end{align}

Constraint \eqref{Eq:1C2} is the only remaining constraint with an explicit dependence on the $x_n$'s. Since $P_{{\rm{B}}n}$ is a function of $x_n$, one could potentially use its inverse to replace $x_n$ with a function of $P_{{\rm{B}}n}$. However, such an inversion function may not exist explicitly (and even if it does, it may
be non-convex). In its stead, we can use a convex upper bound approximation $F$ of the inversion of blocking
probability, so that
\begin{align}\label{Eq:x}
x_n \le F ({P_{{\rm{B}}n}}),\ \forall n\in\mathcal{N}.
\end{align}
Hence, the new convex optimization problem that approximates the original one when $y$ is fixed, is the following:
\begin{alignat}{2}    \label{Eq:minGE3}
\mathop {\min\ } \limits_{\bold P_{\rm{B}}} \sum\limits_{n = 1}^N
[E_{n}^{\rm{L}}(P_{\rm{B}n}) + E_{n}^{\rm{T}}&(P_{\rm{B}n})] \text{ s.t.}\\
\sum\limits_{n = 1}^N {{\alpha _n}F ({P_{{\rm{B}}n}})}  &\le {B^{\max }} - \beta f^{\rm{C}}y\ &&   \label{Eq:21C2} \\
\sum\limits_{n = 1}^N {( {1 - {P_{{\rm{B}}n}}} ) {{\lambda _n}} } &\le {\lambda ^*}   \label{Eq:21C3} \\
P_{{\rm{B}}n}^{\text{min}} \le P_{{\rm{B}}n} &\le 1, &&\forall  n\in\mathcal{N} \label{Eq:21C1}.
\end{alignat}
After solving \eqref{Eq:minGE3}-\eqref{Eq:21C1} and obtaining the $P_{{\rm{B}}n}$'s, we can compute the largest integral $x^*_n$ which achieves a blocking probability equal to or bigger than $P_{{\rm{B}}n}$, for all $n\in\mathcal{N}$.

\begin{algorithm}[h!]
\caption{General Case Approximation for Soft Deadlines (GCASD)}
\begin{algorithmic}[1]  \label{algo:1}

\REQUIRE $\lambda_n, p^{\rm{T}}, p^{\rm{L}}, \alpha _n, K_n, \beta, f^{\rm{C}}, Y, s_j, d_j, q_j, P_j^{\rm{C}}, P^{\rm{G}}_{n,j,k}$, PDFs of $t^{\rm{W}}, t^{\rm{C}}$

\STATE{$cost^*=\infty$}
\FORALL {$a=0,\ldots,Y$}
\STATE {$y=a/Y$}
\STATE {Obtain $\lambda^*$, the upper bound of $\lambda$, by binary search in $[0,\mu^{\rm C}]$} \label{alg:bins}
\STATE {$[P_{\rm{B}}, cost]=$ [solution, objective] of \eqref{Eq:minGE3}-\eqref{Eq:21C1}}
\STATE {$x_{int}=$ max integral $x$ with blocking probabilities $\geq P_{\rm{B}}$}   \label{li:bin1}
\IF {$cost<cost^*$}
\STATE {$x^*=x_{int}; y^*=y; cost^*= cost$}
\ENDIF
\ENDFOR
\RETURN {$x^*,y^*$}
\end{algorithmic}
\end{algorithm}

\textbf{\emph{Complexity Analysis}}: Algorithm GCASD (cf. Algorithm \ref{algo:1}) codifies the solution method described above. Problem \eqref{Eq:minGE3}-\eqref{Eq:21C1} is convex, and can be solved in time $O({\cal L})$, for a polynomial $\cal L$.
Line \ref{li:bin1} takes time $O(N\log K_{max})$ (recall that there are $N$ BSs, and $K_{max}$ is the largest $K_n$).
Hence Algorithm \ref{algo:1} has a running time of $O(Y({\cal L}+\log\frac{\mu^{\rm C}}{\epsilon}+N\log K_{max}))$, where $Y$ is the granularity of $y$, and $O(\log \frac{\mu^{\rm C}}{\epsilon})$ is the binary search cost of line \ref{alg:bins} of the algorithm, in order to get a $\lambda^*$ within $\epsilon$ of the optimal.

\subsection{Approximate Solution for Hard Deadlines}
\label{sec:approximateharddeadlines}

In this subsection, we use a similar approach in order to solve \eqref{Eq:minGCLE}-\eqref{Eq:1CC4}.
Here we decompose the original problem into several subproblems by discretizing both variable $y$ as before, {\em and} $\lambda$. Then, for every possible (fixed) pair $(y,\lambda)$, the non-convex problem \eqref{Eq:minGCLE}-\eqref{Eq:1CC4} can be transformed into a convex optimization problem with $P_{{\rm{B}}n}$ as its decision variables, which can be solved in polynomial time. By calculating the pair $(y^*,\lambda^*)$ whose subproblem achieves minimum average power consumption, integer values $x_n^*$ for the original optimization problem are obtained from $P_{{\rm{B}}n}^*$.

In more detail, we discretize $y\in [0,1]$ by breaking $[0,1]$ into $Y$ equal segments, and then we discretize
$\lambda\in [0, yf^{\rm{C}}/\textstyle\sum_{j = 1}^J P_j^{\rm{C}}q_j]$ by breaking interval $[0, yf^{\rm{C}}/\textstyle\sum_{j = 1}^J P_j^{\rm{C}}q_j]$ into $\Lambda$ equal segments. At iteration $(m,i)$ of this discretization, $y=y^{(m)}$ and $\lambda= \lambda^{(i)}$ are fixed. Then $\Pr[ {{{\tilde t}_{n,j,k}^{\rm{C}}} = t - l}]$ can be calculated directly for any $t$ and $l$, and the original optimization problem~\eqref{Eq:minGCLE}-\eqref{Eq:1CC4} becomes
\begin{alignat}{2}  \label{Eq:minGCLE1}
\mathop {\min } \limits_{\bold{x}}\sum\limits_{n = 1}^N [E_n^{\rm{L}}(x_n) + E_n^{\rm{C}}&(x_n)] \text{ s.t.} \\
\sum\limits_{n = 1}^N {{\alpha _n}{x_n}}  &\le {B^{\max }} -  \beta f^C y^{(m)}\ \ &&  \label{Eq:2CC2} \\
\sum\limits_{n = 1}^N {( {1 - {P_{{\rm{B}}n}}} ) {{\lambda _n}} } &\le \lambda^{(i)} \label{Eq:2CC3} \\
x_n &\in \{0,1,\ldots,K_n\}, &&\forall  n\in\mathcal{N}    \label{Eq:2CC1}
\end{alignat}
This is still a non-convex non-linear integer program, which cannot be solved efficiently. As in Section \ref{sec:systemmodel}, and by using \eqref{Eq:Ineq}-\eqref{Eq:x}, it becomes
\begin{alignat}{2}  \label{Eq:minGCLE2}
\mathop {\min } \limits_{\bold P_{\rm{B}}} \sum\limits_{n = 1}^N [E_n^{\rm{L}}(P_{\rm{B}n}) + &E_n^{\rm{C}}(P_{\rm{B}n})] \text{ s.t.} \\
\sum\limits_{n = 1}^N {{\alpha _n}F ({P_{{\rm{B}}n}})}  &\le {B^{\max }} - \beta f^C y^{(m)}\ \ &&  \label{Eq:3CC2} \\
\sum\limits_{n = 1}^N {( {1 - {P_{{\rm{B}}n}}} ) {{\lambda _n}} } &\le \lambda^{(i)}  \label{Eq:3CC3} \\
P_{{\rm{B}}n}^{\text{min}} \le P_{{\rm{B}}n} &\le 1, &&\forall  n\in\mathcal{N} \label{Eq:3CC4}.
\end{alignat}

Problem \eqref{Eq:minGCLE2}-\eqref{Eq:3CC4} is a convex program and can be solved efficiently. Hence, we can obtain the optimal blocking probabilities $P_{{\rm{B}}n}^*$, corresponding to a pair $(y^{(m)},\lambda^{(i)})$. We can compute the largest integral $x^*_n$ which achieves blocking probabilities no smaller than $P_{{\rm{B}}n}^*$, for all $n\in\mathcal{N}$, by using binary search based on the fact that the $P_{Bn}$'s are decreasing functions of the $x_n$'s. After collecting the solutions for all iterations $(m,i)$, we output the minimum cost one $\bold{x}^*,y^*$.

\begin{algorithm}[h!]
\caption{General Case Approximation for Hard Deadlines (GCAHD)}
\begin{algorithmic}[1]  \label{algo:2}

\REQUIRE $\lambda_n, p^{\rm{T}}, p^{\rm{L}}, \alpha _n, K_n, \beta, f^{\rm{C}}, Y, s_j, d_j, q_j, \Lambda, P_j^{\rm{C}}, P^{\rm{G}}_{n,j,k}$, PDFs of $t^{\rm{W}}, t^{\rm{C}}$

\STATE{$cost^*=\infty, y=0, \lambda=0$}
\WHILE {$y \le 1$}
\WHILE {$\lambda \le  yf^{\rm{C}}/\textstyle\sum_{j = 1}^J P_j^{\rm{C}}q_j$}
\STATE {$[P_{\rm{B}}, cost]=$ [solution, objective] of \eqref{Eq:minGCLE2}-\eqref{Eq:3CC4}}   \label{li:sol}
\STATE {$x_{int}=$ max integral $x$ with blocking probabilities $\geq P_{\rm{B}}$}   \label{li:bin}
\IF {$cost<cost^*$}
\STATE {$x^*=x_{int}; y^*=y; cost^*= cost$}
\ENDIF
\STATE{$\lambda = \lambda+\frac{yf^{\rm{C}}/\textstyle\sum_{j = 1}^J P_j^{\rm{C}}q_j}{\Lambda}$}
\ENDWHILE
\STATE {$y= y+\frac{1}{Y}$}
\ENDWHILE
\RETURN {$x^*,y^*$}
\end{algorithmic}
\end{algorithm}

\textbf{\emph{Complexity Analysis}}: Algorithm GCAHD (cf. Algorithm \ref{algo:2}) codifies the solution method described above. Problem \eqref{Eq:minGCLE2}-\eqref{Eq:3CC4} is convex, and can be solved (line \ref{li:sol})
in time $O({\cal L})$, for a polynomial $\cal L$.
Line \ref{li:bin} takes time $O(N\log K_{max})$ (recall that there are $N$ BSs, and $K_{max}$ is the largest $K_n$). Hence Algorithm \ref{algo:2} has a running time of $O(Y\Lambda({\cal L}+N\log K_{max}))$, where $Y$ and $\Lambda$ are the granularity of $y$ and $\lambda$ respectively.


\section{Task Arrival and Offloading Assumptions}
\label{sec:poissonassumptions}

In the remainder of this paper, we assume that tasks arrive from the MDs at BS $n$ according to a Poisson process with mean arrival rate $\lambda_n$. The Poisson process assumption is commonly made in this type of situation, since the number of mobile devices in a given coverage area is typically quite large, each contributing to a small fraction of the total load \cite{11}. In this case, we can invoke the \emph{insensitivity property} of the Erlang B formula, to compute the probability of blocking at each BS \cite{9}. Note that, typically, the Erlang B result is derived using the $M/M/N/N$ Markovian queue, which assumes exponentially distributed channel upload (i.e., service) times \cite{3}. Due to insensitivity, the result holds for any service time distribution with the same mean. Therefore, the blocking probability for a task arriving at BS $n$ is
\begin{equation}
\label{Eq:PB}
{P_{{\rm{B}}n}} = {\left( {\frac{{{\lambda _n}}}{\mu_n^{\text{W}} }} \right)^{{x_n}}}\frac{1}{{{x_n}!}}{\left[ {\sum\limits_{r = 0}^{{x_n}} {{{\left( {\frac{{{\lambda _n}}}{\mu_n^{\text{W}} }} \right)}^r}\frac{1}{{r!}}} } \right]^{ - 1}}
\end{equation}
where $\mu_n^{\text{W}}$ denotes the mean service rate, which can be calculated by $\mu_n^{\text{W}}= 1/ {{\bar t}_n^{\rm{W}}}$. Function \eqref{Eq:PB} is convex in $x_n$ \cite{4}.

Note that due to the Poisson process task arrival assumption, the channel state sampled by arriving tasks is given by the steady-state equilibrium probability distribution of the Markovian channel at that MD. This follows from the PASTA rule~\cite{10}.

We assume that the aggregate task arrival process at ES is Poisson~\cite{7}, and, therefore, arriving tasks sample the asymptotic equilibrium state distribution of ES. This approximation is justified due to the mixing of arrivals at ES from BSs operating independently. In this case,  ES can be modeled as an $M/G/1$ queue, whose waiting time is given by the random variable $w^{\rm{C}}$. Given $\lambda$ and knowledge of the data upload distribution, the distribution of $w^{\rm{C}}$ can be obtained by numerical inversion of the probability generating function of system waiting time for $M/G/1$ \cite{11}. In this case, the execution time of a task at the ES depends only on which class it belongs to, i.e., $t_{n,j,k}^{\rm{C}}=t_{j}^{\rm{C}}$, for all $n$ and $k$, and $t_j^{\rm{C}} = w^{\rm{C}} + q_j/yf^{\rm{C}}$. Thus, ${\rm{Pr}}[ {t_{n,j,k}^{\rm{W}} + t_{j}^{\rm{C}} \le {d_j}}]$ can be easily obtained.

When applying algorithms GCASD (Algorithm \ref{algo:1}) and GCAHD (Algorithm \ref{algo:2}) in this case, the upper bound $F$ used in problem \eqref{Eq:minGE3}-\eqref{Eq:21C1} and \eqref{Eq:minGCLE2}-\eqref{Eq:3CC4} becomes \cite{8}:
\begin{equation}
x_n \le  \frac{\lambda_n }{\mu_n^{\rm{W}}}(1 - P_{{\rm{B}}n}) + \frac{1}{P_{{\rm{B}}n}},\ \forall n.
\end{equation}

\subsection{Approximation with Soft Deadlines}
\label{sec:poissonapproximatesoftdeadlines}

In this case, problem \eqref{Eq:minGE3}-\eqref{Eq:21C1} becomes:
\begin{alignat}{2}   \label{Eq:minE3}
\mathop {\min\ }\limits_{\bold P_{\rm{B}}} \sum\limits_{n = 1}^N [ E_{n}^{\rm{L}}(P_{{\rm{B}}n}) + E_{n}^{\rm{T}}(P_{{\rm{B}}n})] &\text{ s.t.}  \\
\sum\limits_{n = 1}^N {\alpha _n}({\frac{{{\lambda _n}}}{{{\mu_n ^{\rm{W}}}}}( {1 - {P_{{\rm{B}}n}}} ) + \frac{1}{{{P_{{\rm{B}}n}}}}}) &\le {B^{\max }} - &&\beta {f^{\rm{C}}}y   \label{Eq:5C2} \\
\sum\limits_{n = 1}^N {(1 - {P_{{\rm{B}}n}})} {\lambda _n} &\le {\lambda ^*}   \label{Eq:5C3}  \\
P_{{\rm{B}}n}^{\text{min}} \le P_{{\rm{B}}n} &\le 1, &&\forall  n\in\mathcal{N} \label{Eq:5C1}.
\end{alignat}
Problem \eqref{Eq:minE3}-\eqref{Eq:5C1} is convex, and can be solved in polynomial time. Hence Algorithm \ref{algo:1} can be implemented efficiently.

\subsection{Approximation with Hard Deadlines}
\label{sec:poissonapproximateharddeadlines}

In this case, problem \eqref{Eq:minGCLE2}-\eqref{Eq:3CC4} becomes
\begin{alignat}{2}   \label{Eq:minCLE}
\mathop {\min } \limits_{\bold P_{\rm{B}}} \sum\limits_{n = 1}^N [E_{n}^{\rm{L}}(P_{{\rm{B}}n}) + E_{n}^{\rm{C}}(P_{{\rm{B}}n})]  &\text{ s.t.} \\
\sum\limits_{n = 1}^N {{\alpha _n}({\frac{{{\lambda _n}}}{{{\mu_n ^{\rm{W}}}}}( {1 - {P_{{\rm{B}}n}}} ) + \frac{1}{{{P_{{\rm{B}}n}}}}})}  &\le {B^{\max }} -&& \beta f^C y^{(m)}     \label{Eq:C2} \\
\sum\limits_{n = 1}^N {( {1 - {P_{{\rm{B}}n}}} ) {{\lambda _n}} } &\le \lambda^{(i)}   \label{Eq:C3} \\
P_{{\rm{B}}n}^{\text{min}} \le P_{{\rm{B}}n} &\le 1, &&\forall  n\in\mathcal{N} \label{Eq:C4}.
\end{alignat}
Problem \eqref{Eq:minCLE}-\eqref{Eq:C4} is convex, and can be solved efficiently.

\section{Simulation Results}
\label{sec:simulationresults}

In this section, we present simulation results to demonstrate the performance of our proposed algorithms GCASD (Algorithm \ref{algo:1}) and GCAHD (Algorithm \ref{algo:2}).
We adopt the two-state Gilbert-Elliot channel model~\cite{1}, i.e., the channel states change by following a Markov chain with two states, ``Good'' (G) and ``Bad'' (B). This model is commonly used to characterize the effects of burst noise in wireless channels, where the channel can abruptly transition between good and bad conditions \cite{blazek2018measurement}. The Gilbert-Elliot channel is a difficult one for computation offloading algorithms to deal with compared to those where there is much more correlation in the channel quality as the offloading progresses. Let $B_g$ and $B_b$, respectively, be the data transmission rate when the channel is in the G and B states.  We consider that all channels have the same $B_g$ and $B_b$ values but differ in their state transition probabilities that result in different propagation models.  The transition probabilities for propagation model $k$ in BS $n$ are denoted as ${{P_{n,k}^{\rm{GG}}}},{{P_{n,k}^{\rm{GB}}}}, ~{{P_{n,k}^{\rm{BG}}}}$, and ${{P_{n,k}^{\rm{BB}}}} $. In each time slot, the channel state Markov chain transitions in accordance with these probabilities. Denote $\pi_{n,k}^{\rm{G}}$ and $\pi_{n,k}^{\rm{B}}$, respectively, as the stationary probabilities of a channel in BS $n$ for propagation model $k$ being in the G and B states. Two sets of simulations are performed with set 1 for single class of tasks and set 2 for multiple classes of tasks. Default parameters used in the simulations are summarized in Table~\ref{parameters}.  The parameter settings that we use were taken from the references \cite{Q4,Q5} and \cite{Q1}. These references summarize parameter settings for various types of applications including those that are inherently delay sensitive, such as gaming, face recognition and healthcare use.  We intentionally use a wide range of parameter values based on the referenced ranges so that we can make conclusions that apply in general settings.

\subsection{Simulation set 1: single class of tasks}
\label{sec:simulationresultssingle}

In this subsection, we will assume that all the tasks generated at the MDs have the same data size $s$ and same computation load $q$, i.e., $s_j=s$ and $q_j=q$ for all $j$.  When the channel is in the G state, the transmission rate of the wireless channel allows a task to be uploaded within one time slot; while when the channel is in the B state, the data transmission rate is zero. Since there is only one class of the tasks, subscript $j$ can be dropped from the notation.

Let $t_{n,k}^{\text{W}}$ be the time needed for uploading a task  in BS $n$ with channel model $k$. The probability that one task in BS $n$ with channel model $k$ can be uploaded in $l$ time slots is given as follows
\begin{equation}
\Pr [ {{t_{n,k}^{\rm{W}}} = l} ] = \left\{ {\begin{array}{ll}
\pi_{n,k}^{\rm{G}}, & \mbox{when} \ l = 1\\
\pi_{n,k}^{\rm{B}}{P_{n,k}^{\rm{BB}}}^{l - 2}{{P_{n,k}^{\rm{BG}}}}, & \mbox{when} \ l\ge 2
\end{array}} \right.
\end{equation}
The mean wireless transmission time of a task in BS $n$ uploaded through a channel with propagation model $k$ can be calculated as follows
\begin{equation}
{{\bar t}_{n,k}^{\rm{W}}}=\sum\limits_{l = 1}^{  \infty} l \Pr [ {{t_{n,k}^{\rm{W}}} = l} ] = 1 + \frac{P_{n,k}^{\rm{GB}}}{{{P_{n,k}^{\rm{BG}}}^2 + {P_{n,k}^{\rm{GB}}}{P_{n,k}^{\rm{BG}}}}}.
\end{equation}
Based on this, the mean wireless transmission time of the tasks in BS $n$ is ${{\bar t}_{n}^{\rm{W}}} = \textstyle\sum_{k = 1}^{{I_n}}P^{\rm{G}}_{n,k}{{\bar t}_{n,k}^{\rm{W}}}$, where $P^{\rm{G}}_{n,k}$ is the probability that a task in BS $n$ is uploaded through a channel with propagation model $k$.

With a single class of tasks, the ES server becomes an $M/D/1$ queueing system, $t_{n,j,k}^{\rm{C}}=t^{\rm{C}}$ for all $n$, $j$ and $k$, and the distribution of delay is given by \cite{6}
\begin{equation}
\label{Eq:tB}
\Pr [ {{t^{\rm{C}}} \le \hat t} ] = \left( {1 - \frac{\lambda }{{{\mu ^{\rm{C}}}}}}\right)\sum\limits_{z = 0}^{\lfloor {\hat t{\mu ^{\rm{C}}}} \rfloor } {\frac{{{{[ {{ \lambda }( {\frac{z}{{{\mu ^{\rm{C}}}}} - \hat t} )} ]}^z}}}{{z!}}} {e^{ - \lambda ( {\frac{z}{{{\mu ^{\rm{C}}}}} - \hat t} )}}
\end{equation}
where $\mu^{\rm{C}} = {y{f^{\rm{C}}}}/{q}$.

For comparison, we also run a discrete event simulation (DES) of the system using the $x_n$'s and $y$ solutions obtained from the proposed algorithms to validate our model assumptions, and these solutions are denoted as DESSD and DESHD, respectively, for the soft deadline (SD) and hard deadline (HD) cases.
In addition, we simulate a DES-based OPT scheme for each proposed algorithm as follows. For GCASD, we first obtain all the possible combinations of $x_n$'s under constraint \eqref{Eq:1C4}; for a given combination of $x_n$'s, we can obtain the solution of $y$ based on \eqref{Eq:1C2} and \eqref{Eq:1C5}, and then check if constraint \eqref{Eq:1C6} is satisfied based on the current set of $x_n$'s and $y$.  If not, we go to the next set of $x_n$'s and repeat this procedure. If it is satisfied, we use this set of $x_n$'s and $y$ to run the DES for the system, and then check if \eqref{Eq:1C3} is satisfied. If not, we proceed to the next combination of $x_n$'s and repeat the above procedure. If the constraints are satisfied, we save the obtained average power. After going through all the possible combinations of $x_n$'s, we obtain the minimum average power and the corresponding $x_n$'s and $y$.
For GCAHD, we first obtain all the possible combinations of $x_n$'s under constraint \eqref{Eq:1CC1}; for a given combination of $x_n$'s, we can obtain the solution of $y$ based on \eqref{Eq:1CC2} and \eqref{Eq:1CC4}, and then check if constraint \eqref{Eq:1CC5} is satisfied based on the current set of $x_n$'s and $y$. If not, we go to the next set of $x_n$'s and repeat this procedure. If it is satisfied, we use this set of $x_n$'s and $y$ to run the DES for the system. Then, we save the obtained mean power consumption. After going through all the possible combinations of $x_n$'s, we obtain the minimum average power and the corresponding $x_n$'s and $y$.

\begin{table}[htbp]
\begin{center}
\caption{Default Parameters}
\label{parameters}
\begin{tabular}{|c|c|c|}
\hline
Parameter & Value in set 1 & Value in set 2 \\
\hline
$\tau$ &  \multicolumn{2}{|c|}{1 s}\\ \hline
$p^\text{L}$ &  \multicolumn{2}{|c|}{ 250 mW}\\
\hline
$p^\text{T}$ &  \multicolumn{2}{|c|}{2.5 mW} \\
\hline
$\lambda_n$ & \multicolumn{2}{|c|}{11, 13, 15 tasks/s} \\
\hline
$K_n$ &  \multicolumn{2}{|c|}{15, 15, 20}\\
\hline
$\alpha_n$ &   \multicolumn{2}{|c|}{1, 1, 1 \$}\\
\hline
$\beta$ &  $0.3\times10^{-6}$  \$  & $0.25\times10^{-6}$  \$ \\
\hline
$f^{\rm{C}}$   &  75M cycles/s & 200M cycles/s\\
\hline
$f$ &  1M cycles/s & 2M cycles/s\\
\hline
$B^{\max}$ & 140 \$  & 90 \$\\
\hline
$Bg, B_b$ & 2M, 0 bits per time slot &5M, 1M bits per time slot  \\
\hline
$s_j$  & 2M bits & 5M, 10M, 15M bits \\
\hline
$d_j$ &  4 s & 6, 11, 16 s\\
\hline
$q_j$ &  3M CPU cycles & 10M, 20M, 30M CPU cycles\\
\hline
\end{tabular}
\end{center}
\end{table}

\begin{figure*}[htbp]
\subfigure[Soft deadlines]{
\begin{minipage}[t]{0.48\textwidth}
\centering
\includegraphics[height=78mm, width=88mm]{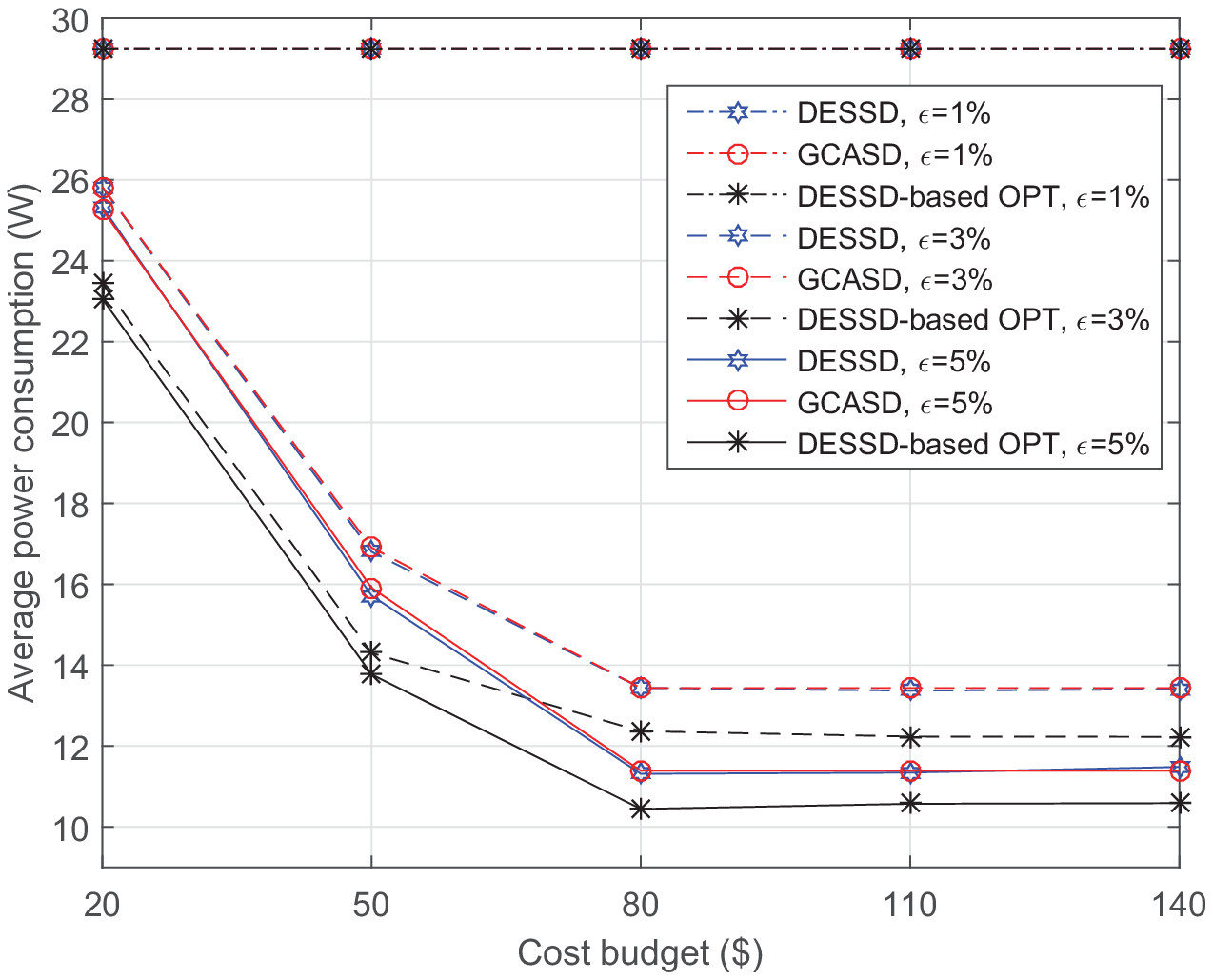}
\label{fig:2}
\end{minipage}
}
\subfigure[Hard deadlines]{
\begin{minipage}[t]{0.48\textwidth}
\centering
\includegraphics[height=78mm, width=88mm]{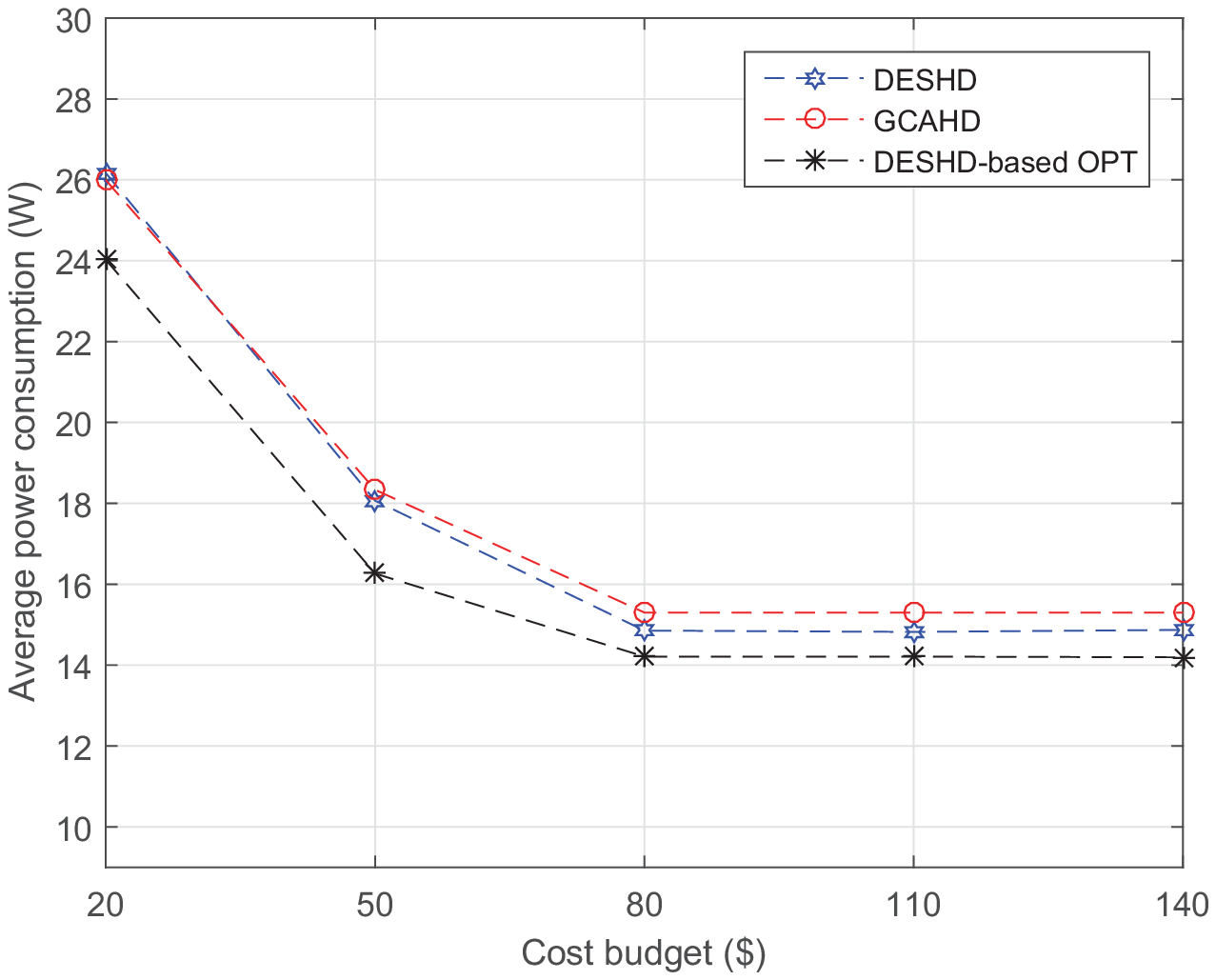}
\label{fig:5}
\end{minipage}
}
\centering
\caption{Average power consumption versus cost budget (Single class of tasks)}
\end{figure*}

\begin{figure*}[htbp]
\subfigure[Soft deadlines]{
\begin{minipage}[t]{0.48\textwidth}
\centering
\includegraphics[height=78mm, width=88mm]{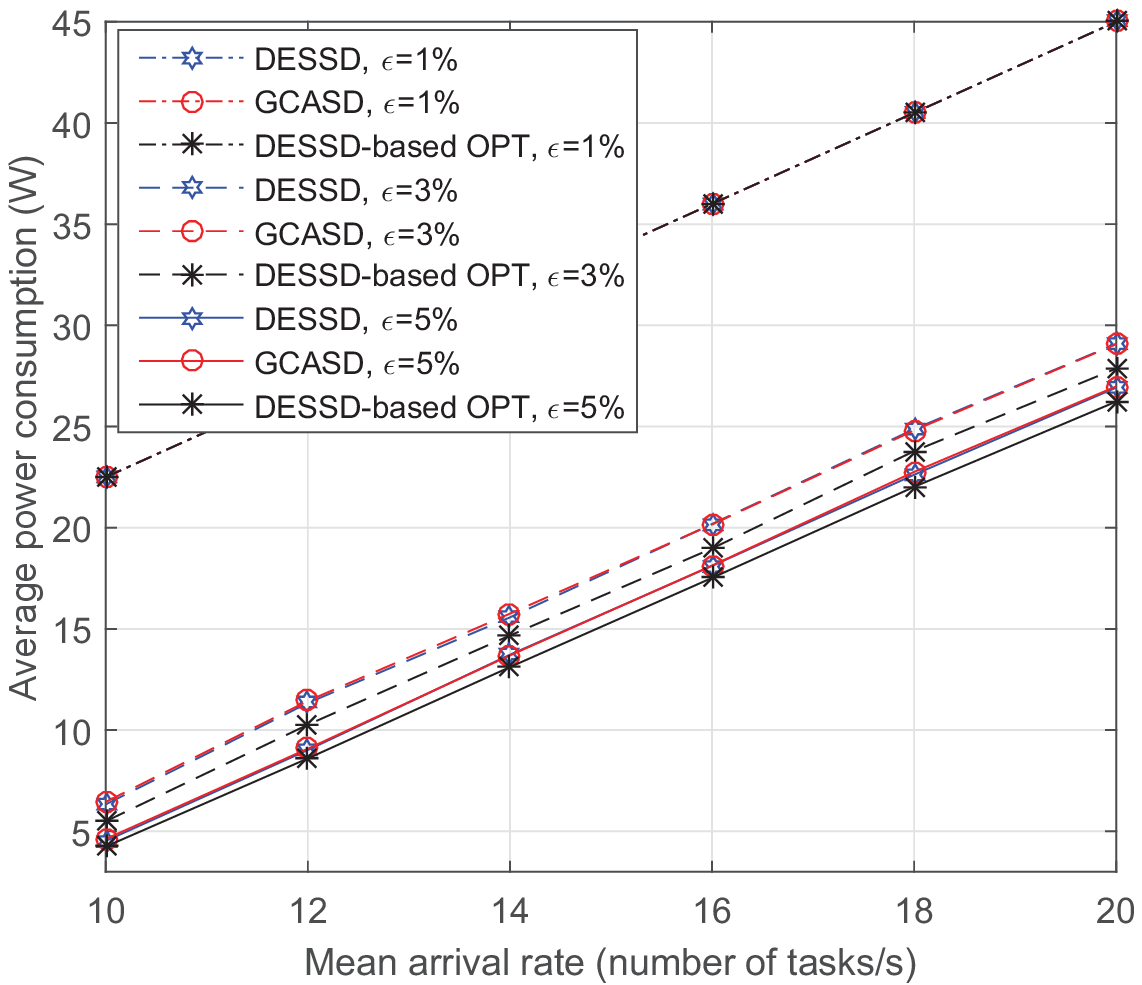}
\label{fig:3}
\end{minipage}
}
\subfigure[Hard deadlines]{
\begin{minipage}[t]{0.48\textwidth}
\centering
\includegraphics[height=78mm, width=88mm]{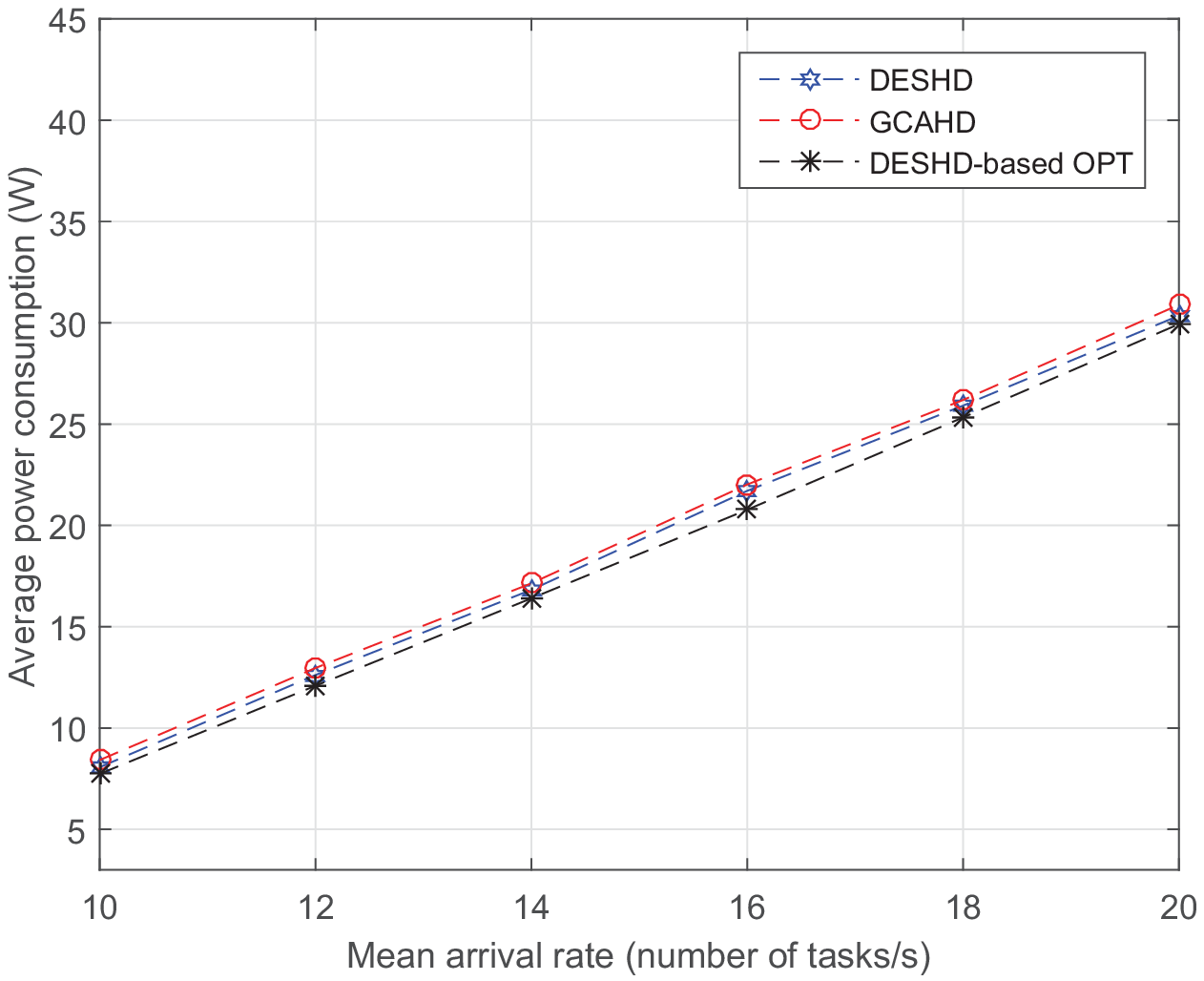}
\label{fig:6}
\end{minipage}
}
\centering
\caption{Average power consumption versus mean arrival rate (Single class of tasks)}
\end{figure*}

In the simulation, we consider a cellular network consisting of 3 BSs.
There are two propagation models at each BS with transition probabilities ${P_{n,1}^{\rm{GG}}}=0.9$, ${P_{n,2}^{\rm{GG}}}=0.7$, ${P_{n,1}^{\rm{BB}}}=0.1$, and ${P_{n,2}^{\rm{BB}}}=0.3$ for $n=1,2,3$.
%
The probabilities of the different channel models in BS 1  are ${P^{\rm{G}}_{1,1}}=0.8$ and ${P^{\rm{G}}_{1,2}}=0.2$; and those in BSs 2 and 3 are ${P^{\rm{G}}_{2,1}}=0.5$, ${P^{\rm{G}}_{2,2}}=0.5$, ${P^{\rm{G}}_{3,1}}=0.2$, and ${P^{\rm{G}}_{3,2}}=0.8$.

\begin{figure*}[htbp]
\subfigure[Soft deadlines]{
\begin{minipage}[t]{0.48\textwidth}
\centering
\includegraphics[height=78mm, width=88mm]{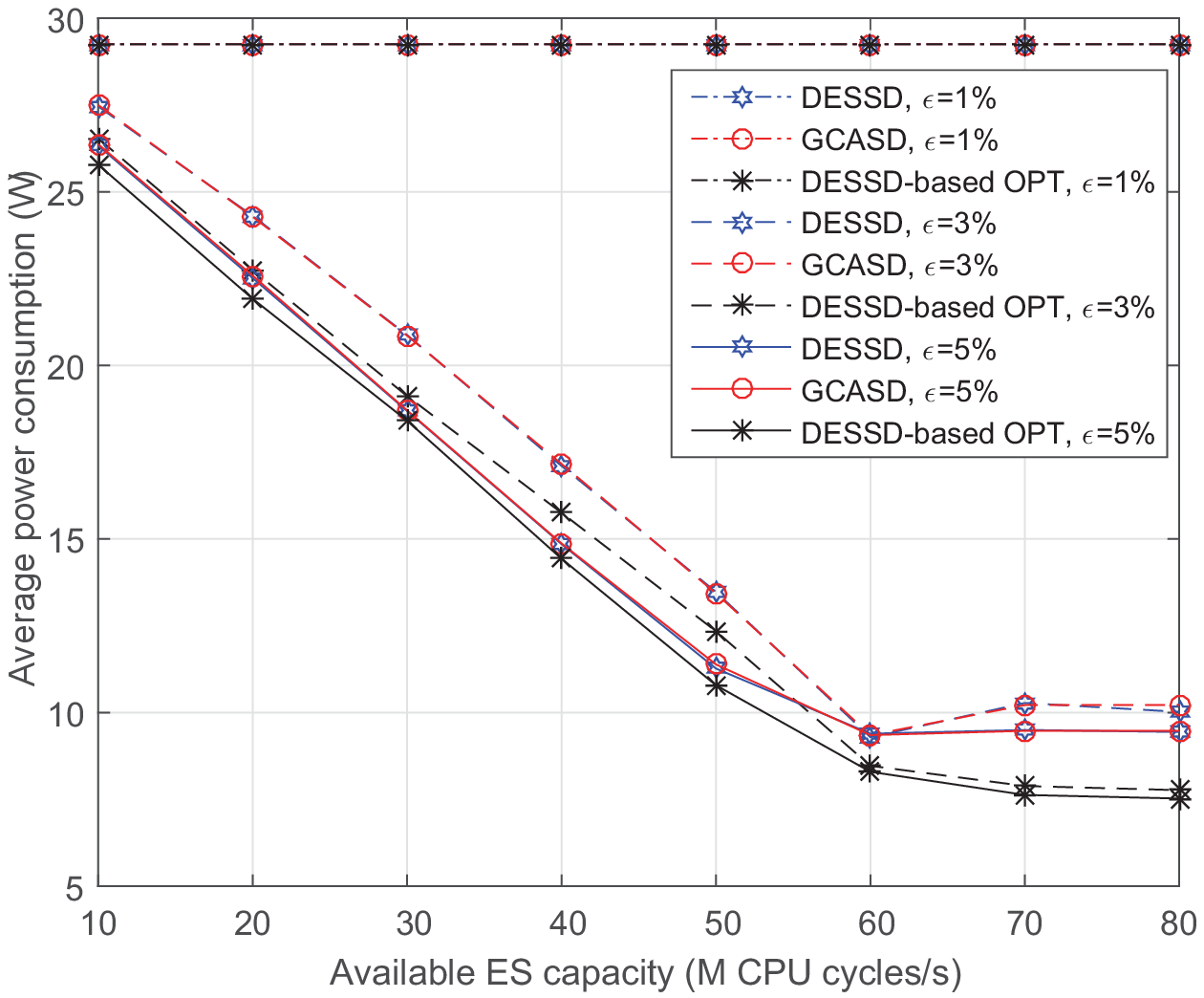}
\label{fig:4}
\end{minipage}
}
\subfigure[Hard deadlines]{
\begin{minipage}[t]{0.48\textwidth}
\centering
\includegraphics[height=78mm, width=88mm]{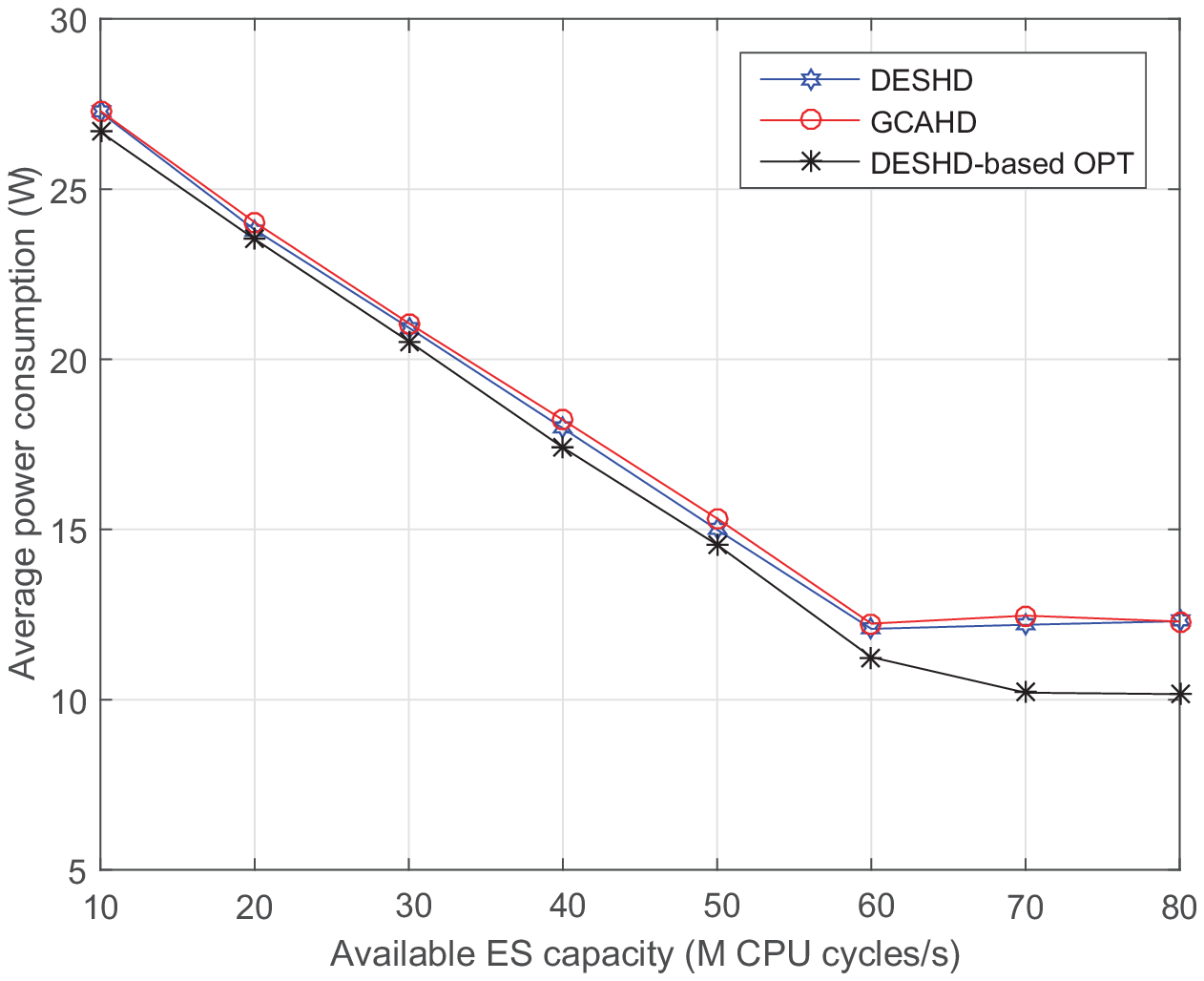}
\label{fig:7}
\end{minipage}
}
\centering
\caption{Average power consumption versus available ES capacity (Single class of tasks)}
\end{figure*}


Figs.~\ref{fig:2} and \ref{fig:5} show the average power consumption of MDs versus $B^{\max}$ for the SD and HD cases, respectively. In Fig.~\ref{fig:2}, when the tolerable violation of latency  $\varepsilon$ is 1\%,  the average power consumption of MDs is a constant for all the solutions. This is because all the tasks are executed locally regardless of the cost budget, since the tight delay constraints cannot be satisfied if a task is offloaded.
When $\varepsilon$ is 3\% or 5\%, some tasks are allowed to be offloaded, and the average power consumption of the MDs decreases with $B^{\max}$ for all the solutions. This happens since,  when the cost budget is small, the optimization is constrained by the cost budget, which limits the number of offloaded tasks; and with the increase of $B^{\max}$, more channel and ES resource is available, leading to more MDs offloading their tasks.
When $B^{\max}$ is large, the budget constraint is loose, and the task offloading completion is mainly affected by the changing wireless transmission conditions. Fig.~\ref{fig:2} also shows that the average MD power consumption decreases with  $\varepsilon$ for all the solutions, since larger $\varepsilon$ makes it easier to meet the latency constraint through offloading, which results in more offloaded tasks and saves power in the MDs.

By comparing the average MD power consumption for $\varepsilon=3\%$ and $\varepsilon=5\%$ in Fig.~\ref{fig:2}, it is seen that the gap is small when the cost budget is small. The gap then increases as the cost budget increases, and finally becomes constant when the cost budget is sufficiently large. When the cost budget is low, the number of channels is small, which forces most tasks to be executed locally, regardless of the value of $\varepsilon$. As the cost budget increases, more channels are available, and the offloading decisions are determined by both $\varepsilon$ and the available channel resources. When the cost budget is sufficiently high, the offloading decisions are mainly determined by the value of $\varepsilon$.
The figure also shows that the average MD power consumption using GCASD is almost the same as using DESSD, which validates the model and approximations used in designing GCASD. The performance of GCASD is also close to DESSD-based OPT, which further shows good performance of the former.

By comparing Figs.~\ref{fig:5} and~\ref{fig:2}, it can be seen that the average MD power consumption for the HD case is slightly higher than that for the SD case with $\varepsilon = 3\%$ and much lower than that for the SD case with $\varepsilon = 1\%$.
For the SD case, when $\varepsilon = 1\%$, the tight (soft) delay constraint forces all the tasks to be executed locally, which results in the highest average power consumption of the MDs;
and the power consumption decreases as $\varepsilon$ increases and more tasks are allowed to be offloaded.
Without having to use CLE, the SD solutions result in lower average MD power consumption than the corresponding HD solutions. However, this is at a price that up to $\varepsilon$ of the tasks do not meet their completion deadlines.
On the other hand, using CLE in the GCAHD only incur slightly higher power consumption of the MDs compared to GCASD when $\varepsilon=3\%$
For the HD case, the total average power consumption of the MDs decreases with $B^{\max}$ when $B^{\max}$ is small and becomes a constant when $B^{\max}$ becomes larger for all schemes, which is the same as that of the SD case with $\varepsilon = 3\%$ and $5\%$.


Figs.~\ref{fig:3} and \ref{fig:6} show the average power consumption versus $\lambda_n$ (same for all BSs) for the SD and HD cases, respectively.
The figures show that the power consumption increases linearly with $\lambda_n$ for all schemes, since both the local execution power and the uploading transmission power are proportional to the mean task arrival rate.
The average MD power consumption using GCAHD is close to that using GCASD with $\varepsilon = 3\%$ but much lower than that using GCASD with $\varepsilon = 1\%$. This demonstrates that the use of CLE in GCAHD is minimized, while always ensuring the HD of the tasks.
Fig.~\ref{fig:3} shows that the performance of GCASD is very close to DESSD and DESSD-based OPT; and Fig.~\ref{fig:6} shows that the performance of GCAHD is very close to DESHD and DESHD-based OPT. These observations are consistent with the ones from Figs.~\ref{fig:2} and~\ref{fig:5}. This further demonstrates the good performance of GCASD and GCAHD and validates the model and approximations used in designing the proposed algorithms.

Figs.~\ref{fig:4} and \ref{fig:7} show the average power consumption of the MDs versus $f^{\rm{C}}$, which is the ES capacity, for the SD and HD cases, respectively. For the SD case with $\varepsilon=1\%$, all tasks are executed locally; and when $\varepsilon=3\%$ and $5\%$, offloading is possible for some tasks, and the number of tasks that can be offloaded increases with the ES capacity, resulting in lower power consumption of the MDs. As the ES capacity is sufficiently high, the average power consumption of MDs becomes a constant, since the offloading decisions are determined by the cost budget which limits the number of wireless channels for uploading tasks.
Note that the slight increase in average power consumption when $f^{\rm{C}}$ is between 60 and 80 is caused by the discretization errors of variable $y$ in algorithms~\ref{algo:1} and~\ref{algo:2}. Increasing the $Y$ values in the algorithms helps reduce the discretization errors but significantly increase the amount of time for running the simulations.
%
%
Comparing the average power consumption of the HD and the SD cases shown in Figs.~\ref{fig:4} and~\ref{fig:7}, we have consistent observations as in previous figures.
%


\subsection{Simulation set 2: multiple classes of tasks}
\label{sec:simulationresultsmultiple}

In this subsection, tasks have multiple classes. The two-state Gilbert-Elliot channels are considered. Let $B_g$ and $B_b$, respectively, be the data transmission rates when a channel is in the G and B states.
%
 Given the channel state transision probabilities, the distribution of wireless transmission time $t^{\rm W}_{n,j,k}$ for uploading a class $j$ task in BS $n$ through a channel with propagation model $k$ can be calculated from~\cite{Arvin}.

At the ES, the system of serving the uploaded tasks becomes an $M/G/1$ queueing system.
Let $B$ be a random variable representing the execution time of the tasks. We have
$\Pr [B=\frac{q_j}{yf^{\rm C}}]=P^{\rm C}_j$, then the probability density function of $B$ can be written as
\begin{align}
f_{B}(\tilde{b}) &= \sum_{j=1}^{J} \Pr\left[B=\frac{q_j}{yf^{\rm C}}\right] \delta\left(\tilde{b}-\frac{q_j}{yf^{\rm C}}\right) \nonumber\\
 &=  \sum_{j=1}^{J} P^{\rm C}_j  \delta\left(\tilde{b}-\frac{q_j}{yf^{\rm C}}\right) ,
\end{align}
and the Laplace-Stieltjes transform of $f_{B}(\tilde{b}) $ is given by
\begin{equation}
g(s)= \sum_{j=1}^J {P^{\rm C}_j}e^{-\frac{q_j}{yf^{\rm C}}s} .
\end{equation}
%
The Laplace-Stieltjes transform of the probability density function of queuing time $w^{\rm{C}}$ is given by the Pollaczek-Khinchine transform~\cite{11} as
\begin{equation}
\label{Eq:WS}
W^*(s)=\frac{(1-\lambda \bar{b})s}{s-\lambda(1-g(s))},
\end{equation}
where $\bar{b}$ is the mean of $B$.
The distribution of $w^{\rm{C}}$ can be obtained by numerical inversion of \eqref{Eq:WS}.

In the simulation, we consider a cellular network consisting of 3 BSs, 3 task classes, and 2 channel propagation models.
The channel state transition probabilities are ${P_{n,1}^{\rm{GG}}}=0.9$, ${P_{n,1}^{\rm{BB}}}=0.1$, ${P_{n,2}^{\rm{GG}}}=0.6$,  and ${P_{n,1}^{\rm{BB}}}=0.4$ for $n=1,2,3$.
The probabilities of accessing channels with different propagation models in BS 1 are ${P^{\rm{G}}_{1,1}}=0.8$ and ${P^{\rm{G}}_{1,2}}=0.2$; those in BSs 2 and 3 are ${P^{\rm{G}}_{2,1}}=0.5$, ${P^{\rm{G}}_{2,2}}=0.5$, ${P^{\rm{G}}_{3,1}}=0.2$, and ${P^{\rm{G}}_{3,2}}=0.8$. The probabilities of a task belonging to different classes are $P^{\rm C}_1=0.6$,  $P^{\rm C}_2=0.3$, and $P^{\rm C}_3=0.1$.

\begin{figure*}[htbp]
\subfigure[Soft deadlines]{
\begin{minipage}[t]{0.48\textwidth}
\centering
\includegraphics[height=78mm, width=88mm]{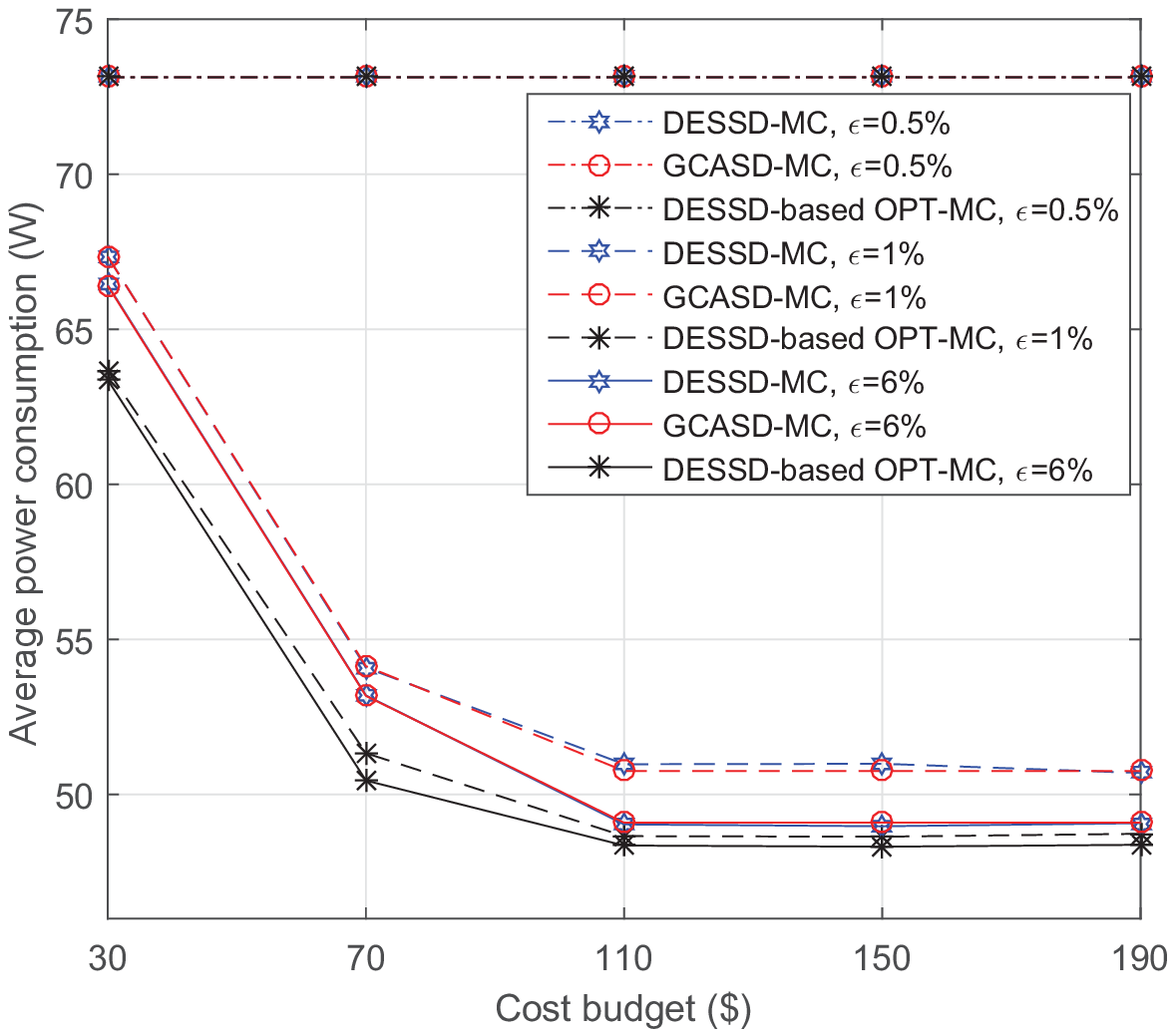}
\label{fig:8}
\end{minipage}
}
\subfigure[Hard deadlines]{
\begin{minipage}[t]{0.48\textwidth}
\centering
\includegraphics[height=78mm, width=88mm]{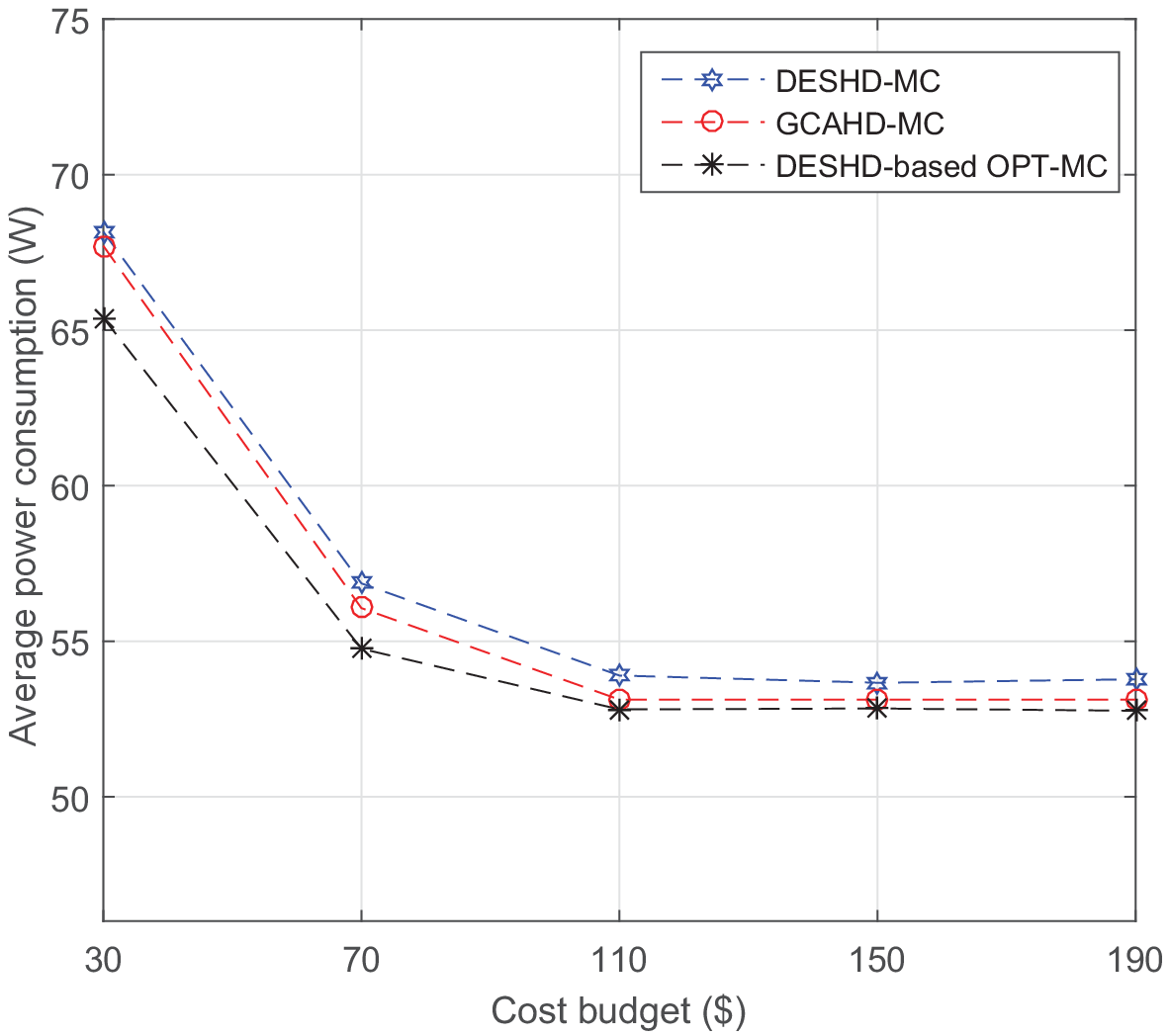}
\label{fig:9}
\end{minipage}
}
\centering
\caption{Average power consumption versus cost budget (Multiple classes of tasks)}
\end{figure*}

\begin{figure*}[htbp]
\subfigure[Soft deadlines]{
\begin{minipage}[t]{0.48\textwidth}
\centering
\includegraphics[height=78mm, width=88mm]{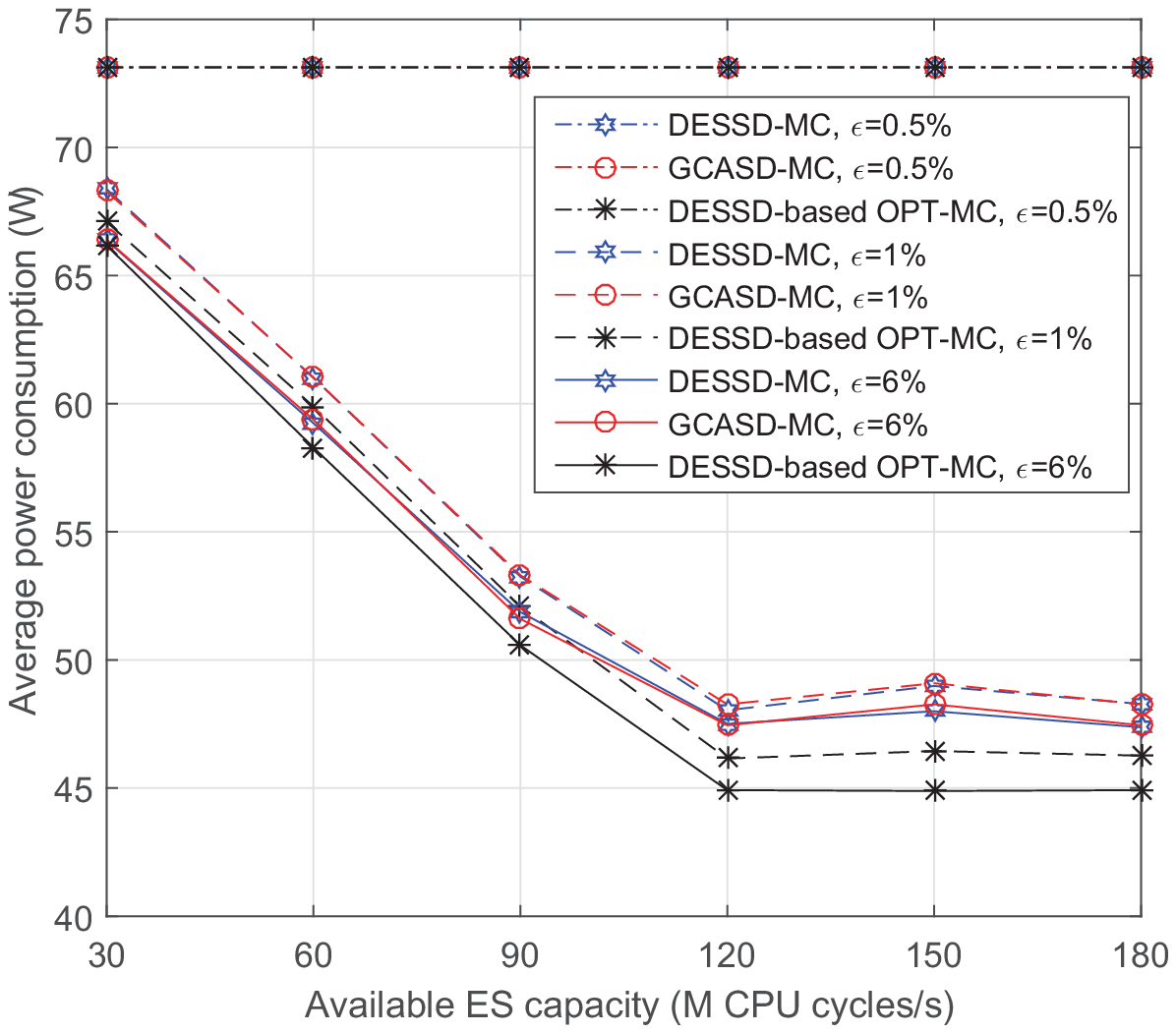}
\label{fig:10}
\end{minipage}
}
\subfigure[Hard deadlines]{
\begin{minipage}[t]{0.48\textwidth}
\centering
\includegraphics[height=78mm, width=88mm]{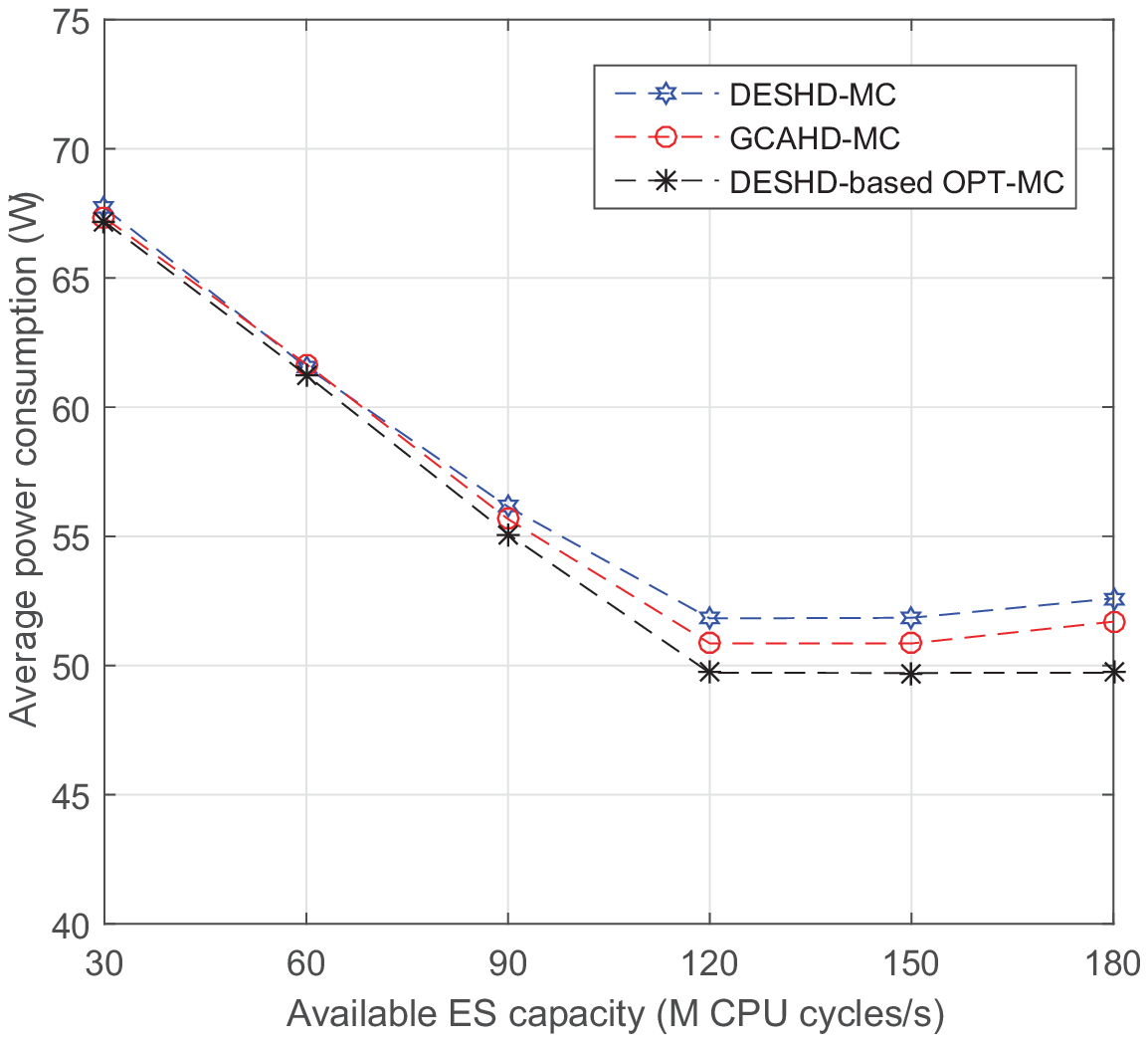}
\label{fig:11}
\end{minipage}
}
\centering
\caption{Average power consumption versus available ES capacity (Multiple classes of tasks)}
\end{figure*}

Figs.~\ref{fig:8} and \ref{fig:9} show the average power consumption of MDs versus $B^{\max}$ for the SD and HD cases, respectively. In Fig.~\ref{fig:8}, when $\varepsilon$ is 0.5\%, all the tasks are executed locally regardless of the cost budget, since offloading cannot satisfy the tight delay constraints.
When $\varepsilon$ is 1\% or 6\%, the average power consumption of MDs decreases with $B^{\max}$ and then becomes a constant.
By comparing the power consumption of the MD in the SD and HD cases, we can see that the average power consumption of MDs for the HD case is slightly higher than that for the SD case with $\varepsilon = 1\%$ and much lower than that for the SD case with $\varepsilon = 0.5\%$.
Figs.~\ref{fig:10} and \ref{fig:11} show the total average power consumption of the MDs versus $f^{\rm{C}}$.
All the results show that our GCASD and GCAHD solutions  achieve the average power consumption performance that is very close to DES-based OPT, and the observations in the multi-class simulations are consistent with the single-class simulations.

\section{Conclusions}
\label{sec:conclusions}
This paper has studied joint wireless network and task service allocation for mobile computation offloading. The objective is to minimize the total average power consumption of MDs for completing the arriving tasks, while satisfying the delay constraints of tasks and the cost budget of the network customer. The formulations presented included both soft and hard task completion time deadlines. The designs were formulated as MINLPs and approximate solutions were obtained by decomposing the formulations into convex subproblems. Simulation results were presented that characterize the performance of the system and show various tradeoffs between task deadline violation, average mobile device power consumption and the cost budget. Results were presented that demonstrate the quality of the proposed solutions, which can achieve close-to-optimum performance over a wide range of system parameters. The optimum allocation were obtained by doing exhaustive search-based discrete event simulations for assigning the wireless channels from each BSs and ES capacity.

\bibliographystyle{IEEEtran}
\bibliography{IEEEabrv,mybib}


\end{document}